\documentclass[sigconf]{acmart}
\copyrightyear{2026}
\acmYear{2026}
\setcopyright{cc}
\setcctype{by}
\acmConference[CCS '26]{Proceedings of the 2026 ACM SIGSAC Conference on Computer and Communications Security}{November 15--19, 2026}{The Hague, Netherlands}
\acmBooktitle{Proceedings of the 2026 ACM SIGSAC Conference on Computer and Communications Security (CCS '26), November 15--19, 2026, The Hague, Netherlands}
\acmDOI{10.1145/3830454.3832577}
\acmISBN{979-8-4007-2871-6/2026/11}
\usepackage{lineno}
\usepackage{tikz}
\usepackage{amsmath}
\usepackage{graphicx}
\usepackage{booktabs}
\usepackage{tikz}
\usepackage{algorithm}
\usepackage{algorithmic}
\usepackage{xspace}
\usepackage{makecell}
\usepackage{varwidth}
\usepackage{tcolorbox}
\tcbuselibrary{skins,breakable}
\usepackage{enumitem}
\usepackage{mdframed}
\usepackage{listings}
\usepackage{amsfonts}
\usepackage{subcaption}
\usepackage{setspace}
\usepackage[table]{xcolor} % enables \cellcolor, \rowcolor in tables
\usepackage{multirow}
\usepackage{listings}
\usepackage{hyperref}
\usepackage{pifont}
\usepackage{seqsplit}

\usepackage{amssymb}
\usepackage[normalem]{ulem}  % normalem 防止它把 \emph 也变成下划线

\newcommand{\bc}{,\allowbreak}
\definecolor{SoftRed}{RGB}{255, 228, 225}
\definecolor{SoftGreen}{RGB}{230, 255, 230}
\definecolor{SoftRed2}{RGB}{255, 102, 102}  
\definecolor{UpGreen}{RGB}{0, 170, 0}
\definecolor{SoftOrange}{RGB}{255, 229, 204}
\definecolor{SoftPurple}{RGB}{224, 224, 248}
\definecolor{SoftBlue}{RGB}{202, 240, 255}
\definecolor{SoftGray}{RGB}{237, 237, 237}
\definecolor{SoftYellow}{RGB}{255, 255, 204}
\colorlet{punct}{red!60!black}
\definecolor{delim}{RGB}{20,105,176}
\colorlet{numb}{magenta!60!black}
\definecolor{background}{RGB}{245,245,245}    % Light gray background
\definecolor{commentcolor}{RGB}{0,128,0}      % Dark green for comments
\definecolor{stringcolor}{RGB}{255,140,0}     % Orange for strings
\definecolor{keywordcolor}{RGB}{0,0,255}      % Blue for keywords
\definecolor{sysbg}{HTML}{F5F0FF}
\definecolor{sysborder}{HTML}{6B4FBB}
\definecolor{userbg}{HTML}{EAF4FF}
\definecolor{userborder}{HTML}{2E6FBB}

\newcommand\tool{\textit{PiMRef}\xspace}
\newcommand\dataset{\textit{SpearMail}\xspace}

\newcommand\llmbenchmarkprofile{{681}\xspace}

\newcommand\llmbenchmark{{14,672}\xspace}

\definecolor{revcolor}{RGB}{0,0,180}

\begin{document}

\title{PiMRef: Deducing Ever-evolving Spear-phishing Emails with Knowledge Base Invariants}

\author{Ruofan Liu}
\email{liu.ruofan16@u.nus.edu}
\orcid{0009-0002-9440-3152}
\affiliation{%
  \institution{National University of Singapore}
  \country{Singapore}
}

\author{Yun Lin}
\authornote{Corresponding author.}
\email{lin\_yun@sjtu.edu.cn}
\orcid{0000-0001-8255-0118}
\affiliation{%
  \institution{Shanghai Jiao Tong University}
  \country{China}
}

\author{Yuxin Wang}
\email{w_yuxin@sjtu.edu.cn}
\orcid{0009-0000-0475-8538}
\affiliation{%
  \institution{Shanghai Jiao Tong University}
  \country{China}
}

\author{Xiwen Teoh}
\email{xiwen.teoh@u.nus.edu}
\orcid{0009-0009-8528-9088}
\affiliation{%
  \institution{National University of Singapore}
  \country{Singapore}
}

\author{Zhenkai Liang}
\email{liangzk@comp.nus.edu.sg}
\orcid{0000-0001-7138-5030}
\affiliation{%
  \institution{National University of Singapore}
  \country{Singapore}
}

\author{Gongshen Liu}
\email{lgshen@sjtu.edu.cn}
\orcid{0000-0001-5194-1570}
\affiliation{%
  \institution{Shanghai Jiao Tong University}
  \country{China}
}

\author{Haojin Zhu}
\email{zhu-hj@sjtu.edu.cn}
\orcid{0000-0001-5079-4556}
\affiliation{%
  \institution{Shanghai Jiao Tong University}
  \country{China}
}

\author{Jin Song Dong}
\email{dcsdjs@nus.edu.sg}
\orcid{0000-0002-8789-2499}
\affiliation{
  \institution{National University of Singapore}
  \country{Singapore}
}

% 问题应对KB构建不透明增加一个专节详述构建流程、错误率、验证方式野外威胁无据
% 补充相关news report/incident report/threat intel citation，或做小规模野外样本分析
% SpearMail评估换成human study或引用更现代的persuasion框架IRB/human study
% 补IRB approval details，说明informed consent情况KB scalability
% 讨论如何扩展到非researcher群体（LinkedIn/公司网站等）
% ORCID数据收集伦理补充数据使用说明，是否符合ORCID ToS

\begin{abstract}
Phishing email is a critical step in the cybercrime kill chain due to the high reachability of victims' email accounts and the low cost of launching phishing campaigns.
The proportion of AI-generated phishing emails peaked at 82.6\% in 2025, and showed a higher click-through rate than manually written ones.
This ever-evolving nature of phishing emails makes traditional rule-based and feature-engineering-based phishing email detectors fight an uphill battle in the cat-and-mouse game of defense and attack.
% Even worse, the emergence of large language models (LLMs) empowers attackers to generate highly convincing emails at even lower costs.

In this work, we show that, large language models (LLMs) can be effectively exploited to generate profile-grounded spear-phishing,
compromising major paradigms of phishing email detectors.
To defend against such LLM-based spear-phishing attacks,
we propose PiMRef,
the first reference-based solution 
to detect ever-evolving phishing emails using knowledge-based invariants, targeting the identity-impersonation attacks that characterize spear-phishing.
Our rationale lies in the fact that convincing phishing emails often include ``disprovable claims'', which contradict certain real-world facts.
Therefore, we reduce the problem of phishing email detection to an identity fact-checking problem on the sender's identity within the email context,
enabling defenses against evolving phishing threats with high accuracy and explainability.
Technically, given an email, 
PiMRef 
(i) discovers the claimed identity of the sender,
(ii) verifies the sender's email domain against a dynamically expandable knowledge base, and
(iii) infers call-to-action instructions that encourage next-step engagement.
The detected contradictory identity facts serve as both alarms and explanations.

Compared to existing baselines such as D-Fence, HelpHed, and ChatSpamDetector, 
\tool reduces the false-positive rate to 0.81\% while maintaining a recall of 90.7\%-93.1\% on conventional phishing benchmarks such as Nazario and PhishPot.
On \dataset, our newly constructed benchmark of 14,672 LLM-generated spear-phishing emails targeting 681 public profiles, 
\tool reaches a recall of 86.4\% without incurring additional false positives.
Furthermore, on 12,289 real-world emails collected from university accounts, spam feeds, and honeypots, \tool achieves a recall of 80.5\%--87.1\% on wild phishing emails, with a median runtime of 0.05 seconds, outperforming state-of-the-art phishing email detectors.
our code is publicly available at \url{https://github.com/code-philia/PhishEmail}.
\end{abstract}



\keywords{phishing detection, spear-phishing, email security, 
large language models}

\maketitle

\section{Introduction}

% 可以和在一起写，大模型 =》lower cost
% 大语言模型的冲击，能力的emergence，更小的代价得到更仿真的内容
% 第一段有焦虑 全是大模型造假的misinformation 怎么verify
% 面向邮件

Large Language Models are a double-edged sword. 
While they help users draft emails and summarize documents, 
the same capability also makes phishing attacks easier to launch and harder to recognize. 
Attackers can now generate high-fidelity, personalized forgeries in seconds with little effort. 
Recent evidence shows that LLMs are already being used to produce spear-phishing emails at scale. 
An industry report finds that 82.6\% of phishing emails contain AI-generated content~\cite{knowbe4_2025}, 
and a controlled human-subject study reports that fully AI-automated spear-phishing achieves a 54\% click-through rate, 4.5$\times$ higher than generic campaigns~\cite{heiding2025evaluating}.
To combat such phishing campaigns, 
existing research on phishing email detection has mainly advanced along two lines: \textit{infrastructure-based authentication} and \textit{content-based detection}. 
% The emergence of Large Language Models has reshaped how humans accomplish day-to-day tasks, such as drafting emails, summarizing documents, and writing code.
% However, the same capabilities that streamline productivity have also lowered the cost of fabricating deceptive content.
% High-fidelity forgeries can be generated within 1 second, without any writing skill or insider knowledge.
% Evidences have shown that attackers are already exploiting LLMs to mount spear-phishing generation at scale.
% Industry reports confirm that 82.6\% of phishing emails contain AI-generated content~\cite{knowbe4_2025}.
% And human recipients are susceptible, a controlled human-subject studies show that fully AI-automated spear-phishing now achieves a 54\% click-through rate, 4.5$\times$ that of generic campaigns~\cite{heiding2025evaluating}.

\textit{Infrastructure-based defenses}, including SPF~\cite{ietf_spf7208}, DKIM~\cite{ietf_dkim6376}, and DMARC~\cite{ietf_dmarc7489}, 
can verify whether an email is sent from an IP address authorized by the claimed sending domain. 
%These mechanisms are effective against domain spoofing by binding the sender infrastructure to the legitimate domain owner. 
However, they do not inspect whether the email content itself is deceptive. 
As a result, an attacker can simply send a phishing email from an attacker-controlled but properly authenticated domain, thereby passing all infrastructure-level checks.

\textit{Content-based defenses}, such as RSpamd~\cite{rspamd}, SpamAssassin~\cite{spamassassin}, and Trend Micro~\cite{trendmicro}, together with machine-learning-based detectors~\cite{ho2019detecting, highprecision, spearphishing, thakur2014catching, identitymailer, duman2016emailprofiler, gascon2018reading, khonji2011mitigation, ma2009detecting, a2011hybrid, khonji2012enhancing, ghosh2023comparison, dfence, helphed, harikrishnan2018machine, shyni2016multi, lee2021catbert, modular, halgavs2020catching}, inspect the email body, header, URL, sender reputation, or other handcrafted and learned features to classify whether an email is phishing. 
These approaches are effective when phishing emails contain recognizable features, such as suspicious wording, abnormal headers, or other statistical patterns that differ from legitimate emails.
However, with LLMs, attackers can easily produce emails with content which is 
(1) \textbf{evasive}, i.e., not exhibit the linguistic features used by traditional content-based detectors and
(2) \textbf{targeting}, i.e., more catering to victims' unique interest to improve the attack success rate.

\begin{figure}[t]
    \centering
    \includegraphics[width=\linewidth]{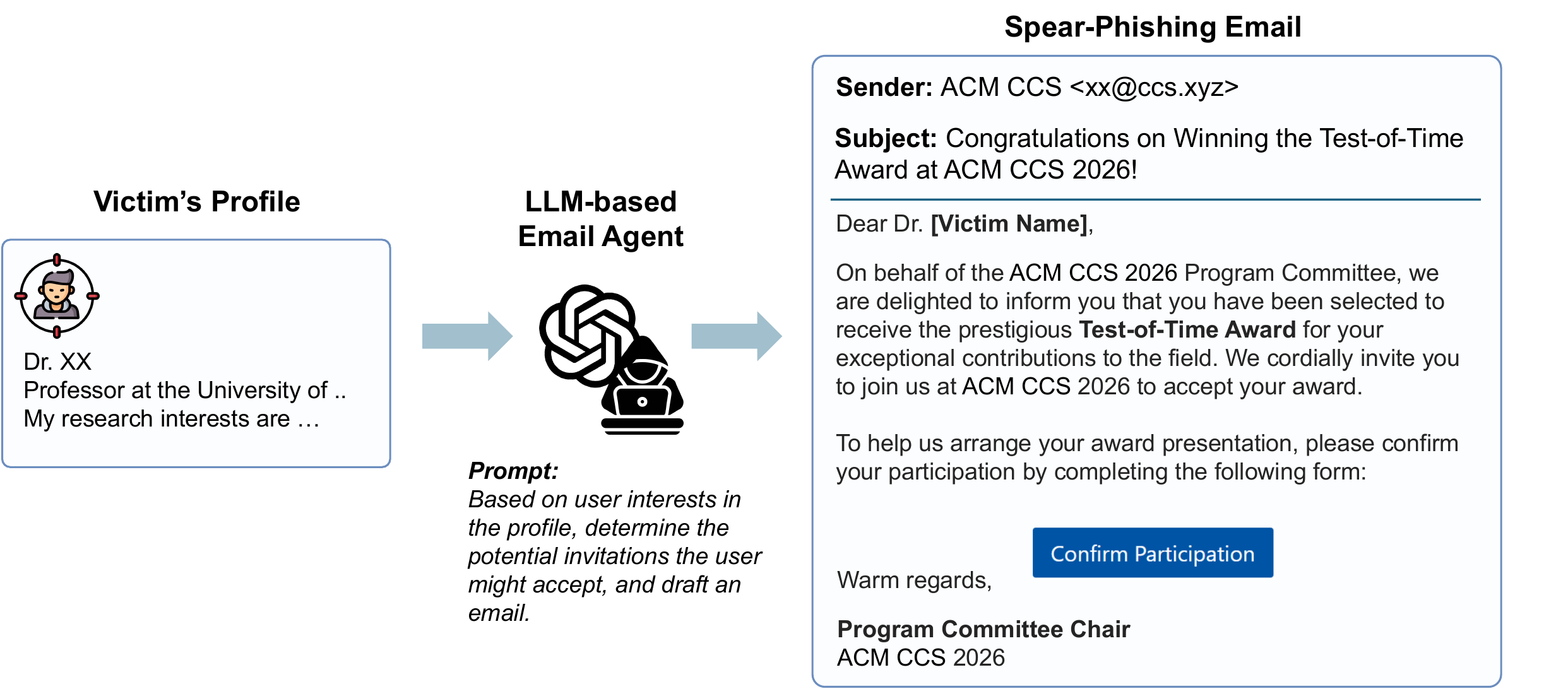}
    \caption{
    Given the victim's profile, an attacker can construct a prompt to generate a spear-phishing email without bypassing LLM safeguards.
    We show that such LLM-generated spear-phishing emails 
    can evade the practically deployable detectors, spanning three major design paradigms (feature-engineering-based, LLM-based, and rule-based)
    (see Section~\ref{exp:closeworld}).
    }
    \label{fig:motivating}
\end{figure}

As illustrated in Figure~\ref{fig:motivating}, given a public victim profile, 
an attacker can infer the victim's interests and 
generate an invitation email tailored to the victim by LLMs, 
then replace the link as a fake website. 
%Such emails can also be sent from attacker-controlled domains that pass SPF, DKIM, and DMARC checks, thereby evading infrastructure-based defenses as well. 
Our experiments confirm that the evaluated detectors struggle under this new threat model (Table~\ref{tab:llm-generated-recall}).
Importantly, this attack does not require explicit jailbreak prompts. 
From the LLM's perspective, the attacker is merely asking it to draft a benign, personalized invitation email, making the generation process difficult to block using jailbreak-oriented safety filters.

% 为什么只用identity fact checking =》cheapest fact
In this work, we present \tool (\textbf{P}h\textbf{i}shing \textbf{M}ail Detection by \textbf{Ref}erence), a reference-based phishing email detector. 
Different from traditional approaches, \tool reformulates phishing detection as a fact-checking problem in the email context. 
The key rationale is that LLM-generated phishing emails can still make claims that are falsified by external facts. 
The contradictions between the claims and the facts provide a more stable signal than historical phishing features. 
Therefore, rather than \textit{inductively} learning what phishing emails used to look like, \tool \textit{deductively} detects phishing emails by verifying the consistency between their claims and relevant real-world references.
However, a naive fact-checking solution is impractical as an email may contain many factual claims, and verifying all of them would be computationally expensive. 
To make reference-based detection practical, \tool focuses on the most inexpensive and effective type of claim, i.e., the \textit{sender identity claim}. 
This claim is highly practical because phishing emails commonly rely on impersonating a trusted sender to establish credibility, 
which can be efficiently verified against a knowledge base that maps legitimate identities to their official email domains. 

To this end, we design \tool to (1) extract identity claims from an email via a named-entity-recognition model over the email header and body, and (2) verify the claimed identity through typo-robust embedding matching against a self-evolving identity knowledge base maintained by an LLM-based web search agent.
As a result, for the phishing email in \autoref{fig:motivating}, \tool generates the following explanation:
``\textit{This email is flagged as phishing because it claims to be from ACM CCS but was sent from a non-official address as \underline{xx@ccs.xyz}}''.

We evaluate the performance of \tool on both conventional phishing benchmarks (i.e., Nazario and PhishPot) and \textit{SpearMail}, a benchmark consisting of \llmbenchmark phishing emails generated from \llmbenchmarkprofile public profiles by LLM.
We compare \tool against
academic baselines (i.e., D-Fence \cite{dfence}, HelpHed \cite{helphed}, and ChatSpamDetector \cite{chatspamdetector}) and
commercial baselines (i.e., RSpamd \cite{rspamd}, SpamAssassin \cite{spamassassin}).
Our results show that \tool reduces the false-positive rate to 0.81\% while maintaining a recall of 90.7\%-93.1\% on conventional phishing benchmarks such as Nazario and PhishPot.
And it significantly boosts recall to 86.4\% on the \textit{SpearMail} dataset.
Additionally, we conducted an open-world study, analyzing 12,289 real-world emails collected over two years.
The study revealed that \tool achieves a precision of 90.29\% and a recall of 80.5\%-100\% on real-world emails, while maintaining an efficient runtime of 0.05 seconds, significantly outperforming the baseline solutions.
In summary, our contributions are as follows:

\begin{itemize}[leftmargin=*]
    \item
    \textbf{Methodology}: To the best of our knowledge, we introduce the first reference-based phishing email detecting methodology to deductively detect and explain phishing emails,
    which is robust to problems of rule degradation and distribution shifts.

    \item
    \textbf{Benchmark}: 
    We show that LLM-empowered spear-phishing emails are realistic threats to commercial and academic anti-phishing solutions, spanning three major design paradigms.
    To this end, we construct the \textit{SpearMail} dataset,
    containing \llmbenchmark spear-phishing emails targeting \llmbenchmarkprofile public profiles, 
    which we use to reveal the vulnerabilities of existing defenses.

    \item
    \textbf{Tool}: We deliver the \tool tool, along with its Outlook plugin version, which can be practically deployed to scan large volumes of incoming emails.
    The code \footnote{\url{https://github.com/code-philia/PhishEmail}} and video \cite{anonymoussite} are available online.

    \item
    \textbf{Evaluation}: We conduct extensive evaluations on both closed-world and open-world experiments, on 12,289 emails over two years,
    showing that \tool can significantly improve the precision and recall over baselines.
\end{itemize}

\section{Related Work}
\noindent\textbf{Phishing Email Detection.}
Employees remain highly susceptible to phishing despite security training~\cite{lain2022phishing, lain2024content, casagrande2023alpha, lain2025url}, making automatic defenses crucial.
Infrastructure-level protocols (SPF, DKIM, DMARC)~\cite{ietf_spf7208, ietf_dkim6376, ietf_dmarc7489} prevent domain-level spoofing but miss targeted or compromised-account attacks~\cite{ashiq2023you, shen2021weak}.
Content-based detection instead inspects the email itself, spanning anomaly detection and hand-crafted rules~\cite{laorden2014study, duman2016emailprofiler, identitymailer, gascon2018reading}, feature-engineered classifiers over body, URLs, and sender reputation~\cite{ho2019detecting, highprecision, spearphishing, dfence, helphed, he2024double}, detectors built on psychological and persuasive cues~\cite{valecha2021phishing, sergeeva2023we, van2019cognitive, Shmalko2022ProfilerDM, Evans2021RAIDERRS}, and industrial spam filters~\cite{rspamd, spamassassin}.
Recent work adds hybrid detectors that combine learned classifiers with LLMs~\cite{ghiuructan2025hybrid}, LLM-based judges that reason over the email content~\cite{chatspamdetector}, and Heiding et al.~\cite{heiding2026evaluating}, who evaluate LLMs on both spear-phishing generation and malicious-intent judging.
All fall into the same inductive paradigm, learning or eliciting what phishing typically looks like.
Such ML-based detectors can be evaded by adversarial perturbations~\cite{spacephish, apruzzese2022mitigating}, and LLMs escalate the threat further.
They can be repurposed to generate phishing content at scale~\cite{roy2024chatbots}, drive web-enabled agents that autonomously harvest personal data for targeted lures~\cite{kim2024llms}, and produce emails that match human experts in click-through rate~\cite{heiding2025evaluating} and rival human-crafted attacks in large-scale organizational deployments~\cite{bethany2025lateral}.
Even reporting pipelines can be undermined through email-tracking-based evasion~\cite{doublydangerous}, so robust content-based detection at the inbox remains a necessary line of defense.

\noindent\textbf{Reference-based Phishing Detection (RBPD).}
RBPD checks a message against a trusted reference set rather than inspecting it in isolation.
In the URL and website setting, RBPD systems maintain logo-domain pairs and flag a page when the displayed brand does not match the actual domain~\cite{phishzoo, visualphishnet, phishpedia, phishintention, phishllm, dynaphish, knowphish, whitenet}, including lightweight client-side variants~\cite{roy2024phishlang}.
Because RBPD grounds its decision in the mismatch between claimed and actual identity rather than in learned phishing patterns, its coverage is bounded by the reference set rather than by the training distribution. 
Detection therefore extends to brands that are rare or absent in phishing corpora, and remains stable as attacker phrasing and formatting drift, provided the reference set is maintained.
\tool\ is, to our knowledge, the first reference-based phishing detector for email, addressing ambiguous identity descriptions, description-domain mapping, and runtime throughput with two encoder-based language models.

\section{Threat Model}\label{sec:threat-model}
Phishing emails in this work are emails whose sender \textit{deceives} the recipient into believing they are someone else, with the intent of compelling the recipient to take a specific action such as replying, clicking an unsafe link, or calling a phone number.
We do not consider general spam such as advertisements or promotions.
Our primary target is \textit{spear-phishing}, where the impersonated identity is personalized to the victim to gain trust.
The same verification applies to generic phishing emails that carry an identity claim, and our evaluation covers both (Section~\ref{exp:closeworld}).

\noindent\textbf{Content-based impersonation.}
We focus on attackers who use attacker-controlled accounts or domains to imitate a sender identity in the email content, without forging or compromising the legitimate domain.
This is a deliberate scope choice, as domain spoofing is an authentication problem best mitigated by strict enforcement of SPF/DKIM/DMARC~\cite{ietf_spf7208, ietf_dkim6376, ietf_dmarc7489}.
An attacker who spoofs the exact official domain of the claimed identity produces a consistent identity-domain pair and raises no alarm in \tool, but such an email fails SPF/DKIM alignment at the receiving server and is rejected or quarantined under a DMARC enforcement policy before content analysis applies.
\tool therefore assumes sender authentication as a deployed lower layer and detects the complementary class of attacks that pass authentication legitimately.
Business Email Compromise likewise involves compromised or domain-consistent accounts and requires signals beyond content-only analysis, such as sender-history anomalies and workflow controls~\cite{travelers_bec,ho2019detecting,highprecision}.
In our observations on PhishPot~\cite{phishpot}, an open phishing-email repository, \textbf{90\%} of phishing emails fall into this content-based setting and only \textbf{10\%} forge the sender address.
Content-based impersonation is also operationally cheaper, requiring only email generation rather than compromising any infrastructure.

\noindent\textbf{Non-triviality.}
We consider a phishing email non-trivial when the attacker both instructs the victim to take a specific action and claims an identity, in the header or the content, to establish trust.
Spear-phishing emails usually satisfy this.
An email claiming to be UPS~\cite{ups} and saying ``\textit{Track your parcel here to ensure delivery}'' is non-trivial, whereas ``\textit{Hi, are you available?}'' is trivial because it specifies no identity.
Attackers may further use web crawlers and LLMs to craft the title, sender identity, and body according to their malicious intent.

\begin{figure*}[ht]
   \centering
   \includegraphics[width=\linewidth]{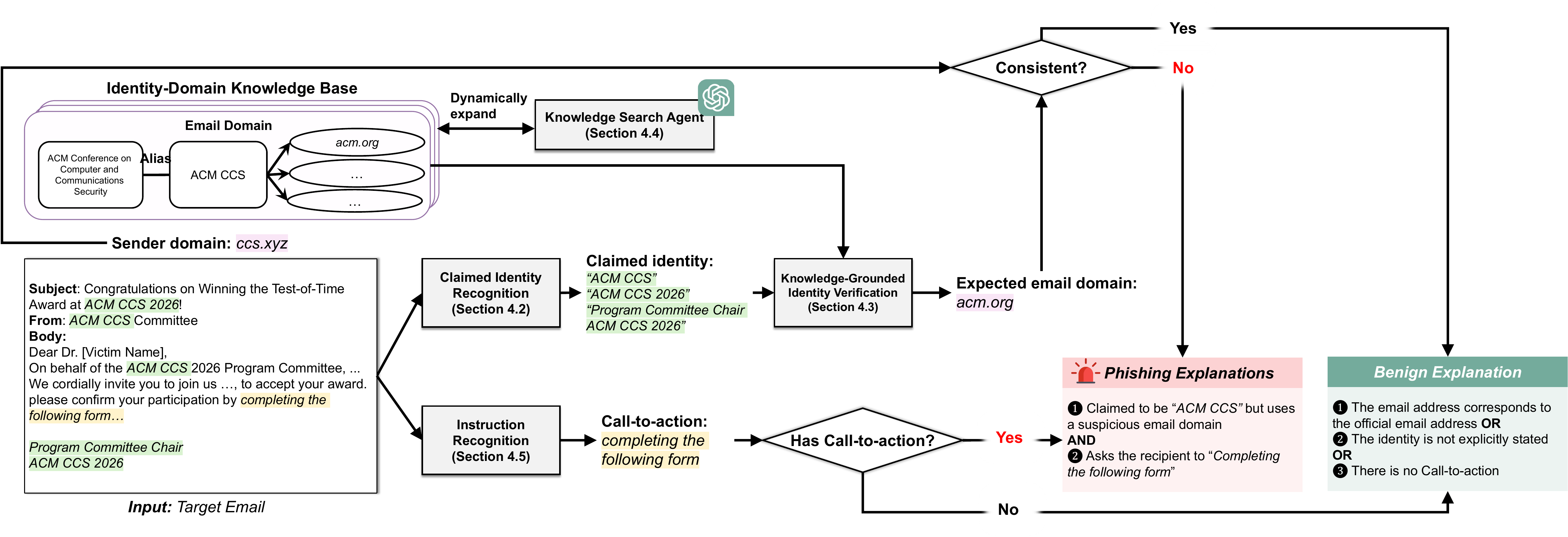}
   \caption{
   Overview of \tool.
   The \textit{Claimed Identity Recognition} module first extracts the phrases claiming the identity.
   The \textit{Knowledge-Grounded Identity Verification} module then converts the identity-claiming phrases into their expected email domains (e.g., \textit{acm.org}),
   based on a predefined \textit{Identity-Domain Knowledge Base}.
   Finally, the \textit{Instruction Recognition} module extracts the phrases of call-to-action instruction in the email.
   \tool reports the phishing alert if the actual email domain is different from all of the expected email domains, and the email has call-to-action instructions.
   }
   \label{fig:overview}
\end{figure*}

\section{Approach}

\noindent\textbf{Overview.}
\autoref{fig:overview} presents the workflow of \tool,
which takes an email as input, and outputs a phishing alert along with its counterfactual explanation if the email is phishing.
\tool consists of three modules, i.e., claimed identity recognition, knowledge-grounded identity verification, and a call-to-action instruction recognition module.
\begin{itemize}[leftmargin=*]
  \item \textbf{Claimed Identity Recognition} (Section \ref{app:identity-recog}):
  This module parses the email subject, sender name, and email body as input to infer the claimed sender identities $\textit{ID}_{rec}$.
  \item \textbf{Knowledge-Grounded Identity Verification} (Section \ref{app:identity-matching}):
  This module maps the predicted identities $\textit{ID}_{rec}$ to their legitimate official email domains $\mathcal{D}_{rec}$, using an identity-domain knowledge base.
  \item \textbf{Instruction Recognition} (Section \ref{app:ins-recog}):
  This module outputs the set of phrases of call-to-action instructions $\textit{inst}$, if any.
\end{itemize}
Then, we verify the consistency between the expected official email domains $\mathcal{D}_{rec}$ and the domain of the sender's email address in the target email.
Specifically, a phishing alert is raised if
(i) the actual email domain $d \notin \mathcal{D}_{rec}$ (i.e., $d$ is not a legitimate domain of any claimed identity) and
(ii) $\textit{inst} \neq \emptyset$ (i.e., the email contains instructions for next-step engagement).

\subsection{Email Pre-processing}
We first render the email body (including HTML content and common attachments) into PDF format, 
on which we run OCR \cite{paddleocr} to recognize the textual content. 
This design makes our pipeline effectively immune to common text-salting attacks \cite{text-salt} 
(e.g., zero-width characters or non-printable characters) that target tokenization at the raw text level.
Then we tokenize each preprocessed email 
$m = \{\textit{name}, \textit{subject}, \textit{body}\}$ 
into sequences of subword tokens using a standard tokenizer~\cite{bert}. 
Formally, 
$\textit{name} = \langle t_{i_1}, t_{i_2}, \dots \rangle$ denotes the sender name, 
$\textit{subject} = \langle t_{j_1}, t_{j_2}, \dots \rangle$ denotes the email subject, and 
$\textit{body} = \langle t_{k_1}, t_{k_2}, \dots \rangle$ denotes the message body.

\subsection{Claimed Identity Recognition}\label{app:identity-recog}

%Phishing emails often make false claims, such as ``this email is sent from Netflix'', ``the account has been outdated'', ``the mailbox is almost full''.
%Therefore, by debunking these false claims, one can deduct the email's legitimacy.
%However, not all claims are easily ``verifiable'', as we do not have access to the victim's actual accounts.
%In this paper, we treat the sender’s identity as a verifiable claim, one that can be cross-referenced with official information.

In this step, we infer all token spans in $m$ which indicate the claimed identities of the sender.
For example, in \autoref{fig:motivating},
we report terms such as \textit{``ACM CCS 2026''}.

\noindent\textbf{Naive Solution and its Practical Challenge.}
Prompt engineering a state-of-the-art LLM \cite{openai2023chatgpt} is common practice for general NLP problems, but it scales poorly here.
Autoregressive decoding makes per-email inference slow, and paying per API call for every email is costly, so the naive solution is impractical for an organization parsing tens of thousands of emails a day.

%Recognizing the sender's identity is a non-trivial task.
%The first challenge arises from the various contexts in which an identity can appear.
%It may be located in different \textit{positions}, such as the sender's name, the email's closing signature, or embedded within the email's content (e.g., Thank you for choosing our ``Paypal'' service).
%Additionally, the identity can be presented in multiple \textit{forms}, including an organization name (e.g., ``PayPal'') or a specific role (e.g., ``Admin'').
%These variations make it challenging to encapsulate all such cases under a unified ``identity'' category.
%
%Another concern is inference efficiency.
%One straightforward solution is to outsource this task to decoder-based language models \cite{anthropic2023claude, openai2023chatgpt, google2023gemini, google2022palm, meta2023llama2, meta2022opt, mistral2023mistral7b} such as GPT.
%However, the auto-regressive generation process of these models leads to an average processing time of 3 seconds per email, making them unsuitable for real-time detection systems.

\noindent\textbf{Our Solution.}
We propose an efficient and lightweight solution to recognize the identity of the sender.
We adopt an encoder-based solution which is smaller in size (e.g., 340M parameters),
but can process all the tokens in an email in parallel,
empirically incurring the average runtime overhead of only 0.02 seconds (see Section~\ref{exp:closeworld} for more details).
Specifically, we reframe the identity recognition task as a Named Entity Recognition (NER) problem \cite{nersurvey}.
%Since the NER model is encoder-based rather than decoder-based, the inference can be parallelized efficiently, with an average runtime of 0.02 seconds.

% \begin{figure}
%   \centering
%   \includegraphics[scale=0.25]{figures/ner-model.pdf}
%   \caption{The application of NER model to infer the token type (i.e., BE for \textit{beginning of an entity}, IE for \textit{inside of an entity}, and O for \textit{outside entity})}\label{fig:ner-model}
% \end{figure}

\begin{figure}[t]
     \centering
     \includegraphics[width=\linewidth]{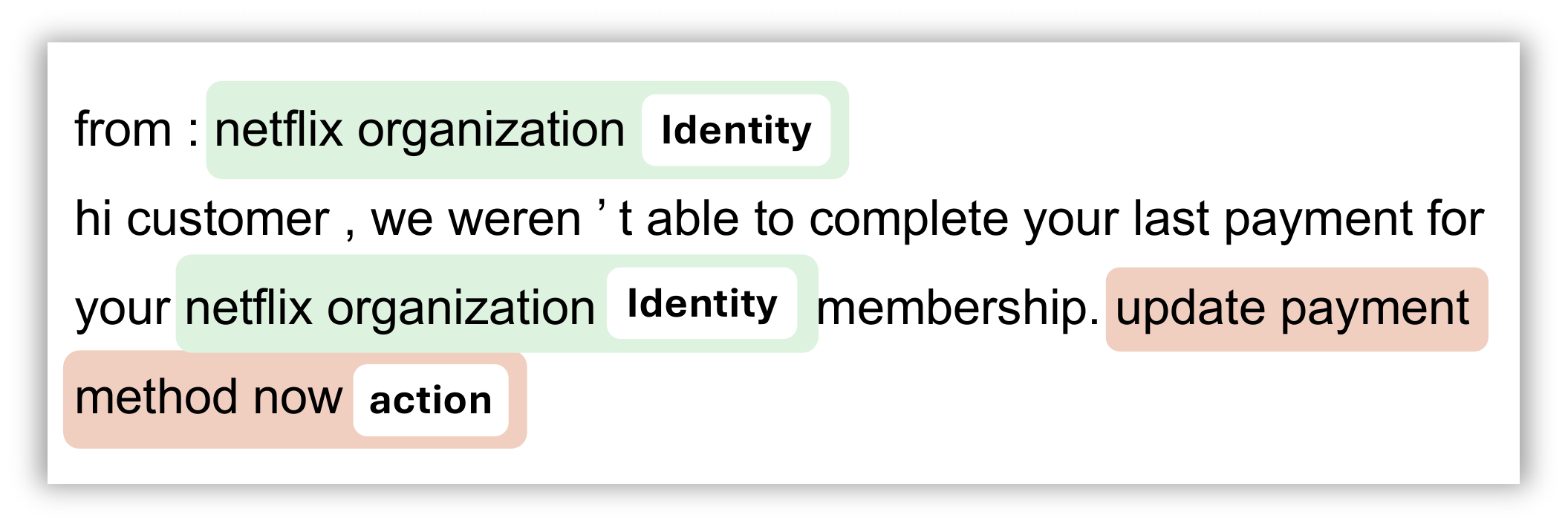}
     \caption{Example of NER training samples.}
     \label{fig:ner_eg}
\end{figure}

\noindent\textbf{Design of Model and Dataset.}
Overall, our NER model takes a sequence of tokens as input,
and assigns each token with one of the following classes, i.e.,
\textit{BE} (i.e., the beginning of entity),
\textit{IE} (i.e., the inside of an entity), and
\textit{O} (i.e., the outside),
following the practice of training any NER models.
In this work, we prepare a dataset of 2,086 emails with labeled entities (from the Nazario and Enron corpora, see Section~\ref{exp:closeworld})
as shown in \autoref{fig:ner_eg},
with the state-of-the-art labeling tool, Label Studio \cite{labelstudio}.
%This paragraph is then analyzed by a trained Named Entity Recognition (NER) model, which tokenizes the text and assigns each token to one of five predefined classes: B-identity (beginning of an identity), I-identity (inside an identity), B-action (beginning of a call-to-action phrase), I-action (inside a call-to-action phrase), and O (outside any class).
% Note that, an enterprise user can receive both external and internal spear-phishing emails.
% Therefore, when we label the tokens of identity entities,
% we additionally consider the phrases indicating the relation between the sender and the recipients,
% such as the \textit{Colleague} in the body as \textit{``Dear Colleague, ...''}.
% Those phrases are useful to infer the internal entities, i.e., the sender claiming to be with an internal email account.
Then, we use the focal loss \cite{focalloss} in \autoref{eq:focal-loss} to train our model.
Here, $i$ indexes each token in the email, and $p_{y_i}$ denotes the predicted probability of the ground-truth class for token $i$.
Focal loss is well-suited for class imbalance tasks,
where negative tokens (O class) outnumber positive tokens (I or B classes).

\begin{equation}\label{eq:focal-loss}
\mathcal{L}_{\text{Focal}} = -\frac{1}{N} \sum_{i=t_1}^{i=t_N} (1 - p_{y_i})^\gamma \, \log(p_{y_i})
\end{equation}

Finally, we augment the training dataset by mutating the phrases of identities (with character-level perturbation) to improve model robustness.

%\linyun{TODO: Add a loss function formula here.}

%For the identity class, we consider not only brand names but also the potential relation of the sender to the recipient (e.g., colleague, friend, manager).

\subsection{Knowledge-Grounded Identity Verification}\label{app:identity-matching}

\begin{figure}[t]
    \centering
    \includegraphics[width=\linewidth]{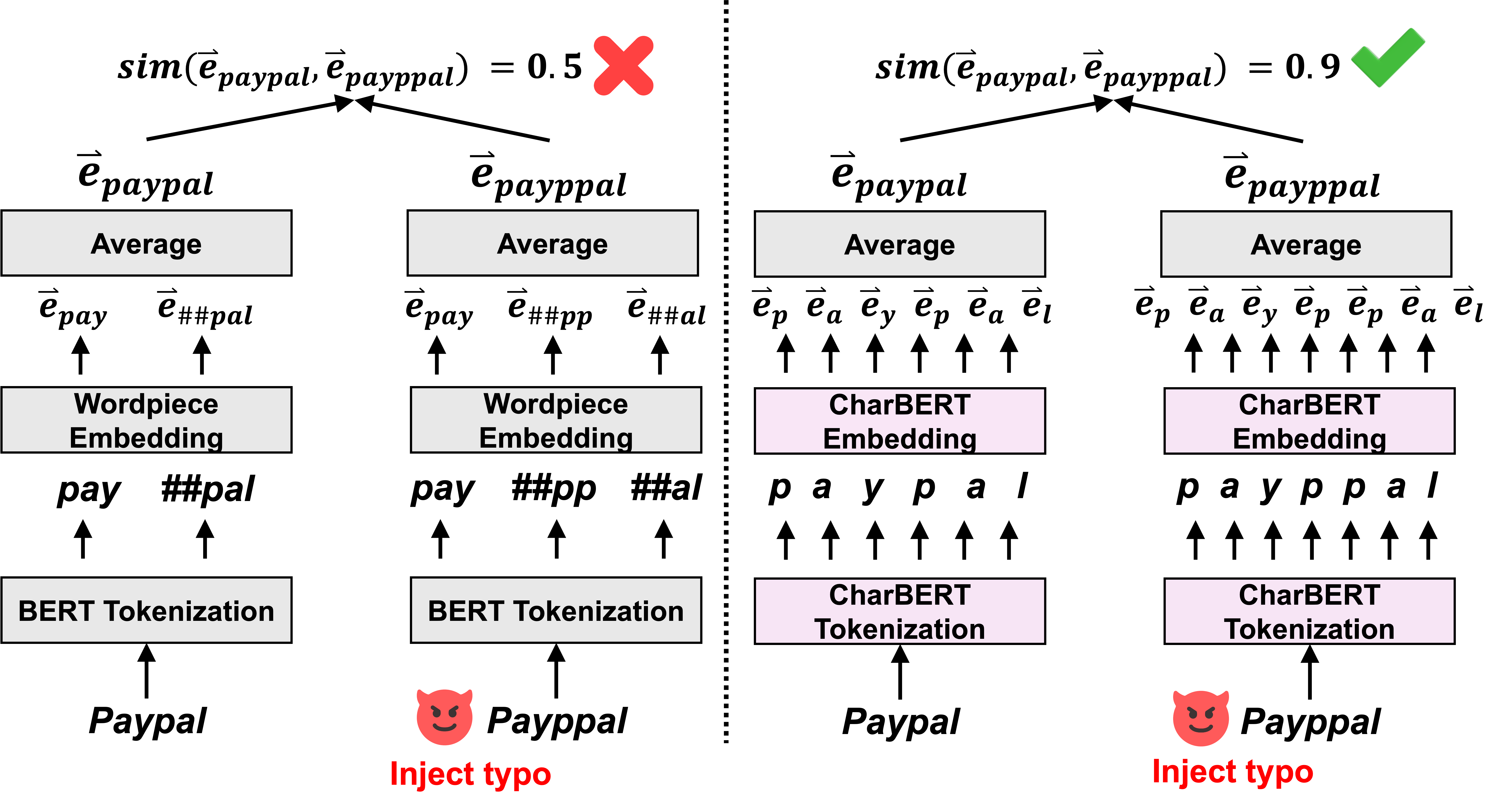}
    \caption{Comparison between BERT (LHS) and CharacterBERT (RHS) when encountering the typo injection attack.
    }
    \label{fig:typo_eg}
\end{figure}

\noindent\textbf{Problem Statement.}
We maintain a knowledge base $\mathit{KB} = \{\mathit{kb}_1\bc \mathit{kb}_2\bc \dots\}$, where each entry $\textit{kb}_i = \langle \textit{id}, d \rangle$ maps an identity $\textit{id} \in \textit{ID}$ to one of its legitimate email domains $d \in \mathcal{D}$, so an identity may appear in multiple entries, and $\textit{ID}$ and $\mathcal{D}$ denote all known identities and all their legitimate domains.
Given the identity-claiming phrases $\textit{ID}_{rec} = \{\textit{id}_1, \textit{id}_2, \dots\}$ extracted in Section~\ref{app:identity-recog}, our task is to retrieve
$\mathcal{D}_{rec} = \{\, d \mid \langle \textit{id}, d \rangle \in \textit{KB},\ \textit{id}\ \text{matches some}\ \textit{id}_j \in \textit{ID}_{rec} \,\} \subseteq \mathcal{D}$,
the set of legitimate domains associated with $\textit{ID}_{rec}$.

\noindent\textbf{Technical Challenge.}
Two difficulties arise.
First, an attacker may fake an \textit{external} identity, an organization outside the recipient's own, or an \textit{internal} one, and internal claims often name no specific identity at all (e.g., ``\textit{Dear colleague, ...}'') yet still require legitimate domains for validation.
Second, exact matching is too restrictive because identity names carry intentional or unintentional typos~\cite{modular}, while edit distance captures string overlap but not semantics, as ``payppall'' is a typo of ``paypal'' whereas the comparably distant ``payday'' is a different word.

\noindent\textbf{Our Solution.}\label{app:external-matching}
We learn an embedding model $f(\cdot): \textit{ID} \rightarrow \mathbb{R}^k$ from identity phrases in natural language to a $k$-dimensional space, projecting semantically similar phrases closer than dissimilar ones.
We adopt CharacterBERT~\cite{characterbert, characterbert-typo}, which tokenizes at the character level rather than the token level, so unaffected characters remain intact under typos.
\autoref{fig:typo_eg} illustrates this, where BERT's tokenization of ``PayPal'' is significantly altered by typos while CharacterBERT preserves the tokenization of the unaffected characters and yields more robust embeddings.
We pre-train on the dataset of~\cite{characterbert-typo} with the loss below, whose first term is a retrieval loss enforcing that the query $q$ stays close to its true neighbor set $p^+$ within the candidate set $\mathcal{P}$, and whose second term is an auxiliary KL divergence requiring a typo-ed $q'$ to yield the same prediction as $q$.

\begin{align}
\scriptsize
\label{eq:kl-loss}
\mathcal{L} =\;
& \underbrace{
   -\log 
   \frac{e^{\bigl(f(q)^\top f(p^+)\bigr)}}
        {\displaystyle\sum_{p \in \mathcal{P}} e^{\bigl(f(q)^\top f(p)\bigr)}}
}_{\mathcal{L}_{\mathrm{Retrieval}}}
+ \underbrace{D_{KL}\Bigl( 
f(q')^\top f(\mathcal{P}) ||
f(q)^\top f(\mathcal{P})
\Bigr)}_{\mathcal{L}_{\text{KL}}}
\end{align}

This resolves both challenges together.
Identity phrases such as \textit{PayPal Inc} or \textit{Colleague} map to knowledge base identities such as \textit{Paypal} or the special identity \textit{Internal} that we introduce for internal claims.
For an external identity we verify that the \textbf{official email domain} is consistent with the \textbf{sender's email domain}; for an internal one we verify that the \textbf{sender's} and the \textbf{recipient's} email domains agree, confirming they belong to the same organization.
The character-level architecture in turn keeps the similarity score robust to intentional or unintentional perturbation of identity phrases.

\noindent\textbf{Subdomain Handling.}
To accommodate legitimate subdomain usage, \tool canonicalizes both the sender domain and the KB domains to their registrable domain, i.e., effective top-level domain plus one label (eTLD+1), before comparison, so a sender domain such as \texttt{cs.cmu.edu} is consistent with the KB entry \texttt{cmu.edu}.
This avoids falsely flagging legitimate emails sent from departmental, regional, or service-specific subdomains of an official organization domain.
\subsection{Knowledge Base Evolution} \label{app:knowledge-construction}

\begin{figure}[h]
    \centering
    \includegraphics[width=\linewidth]{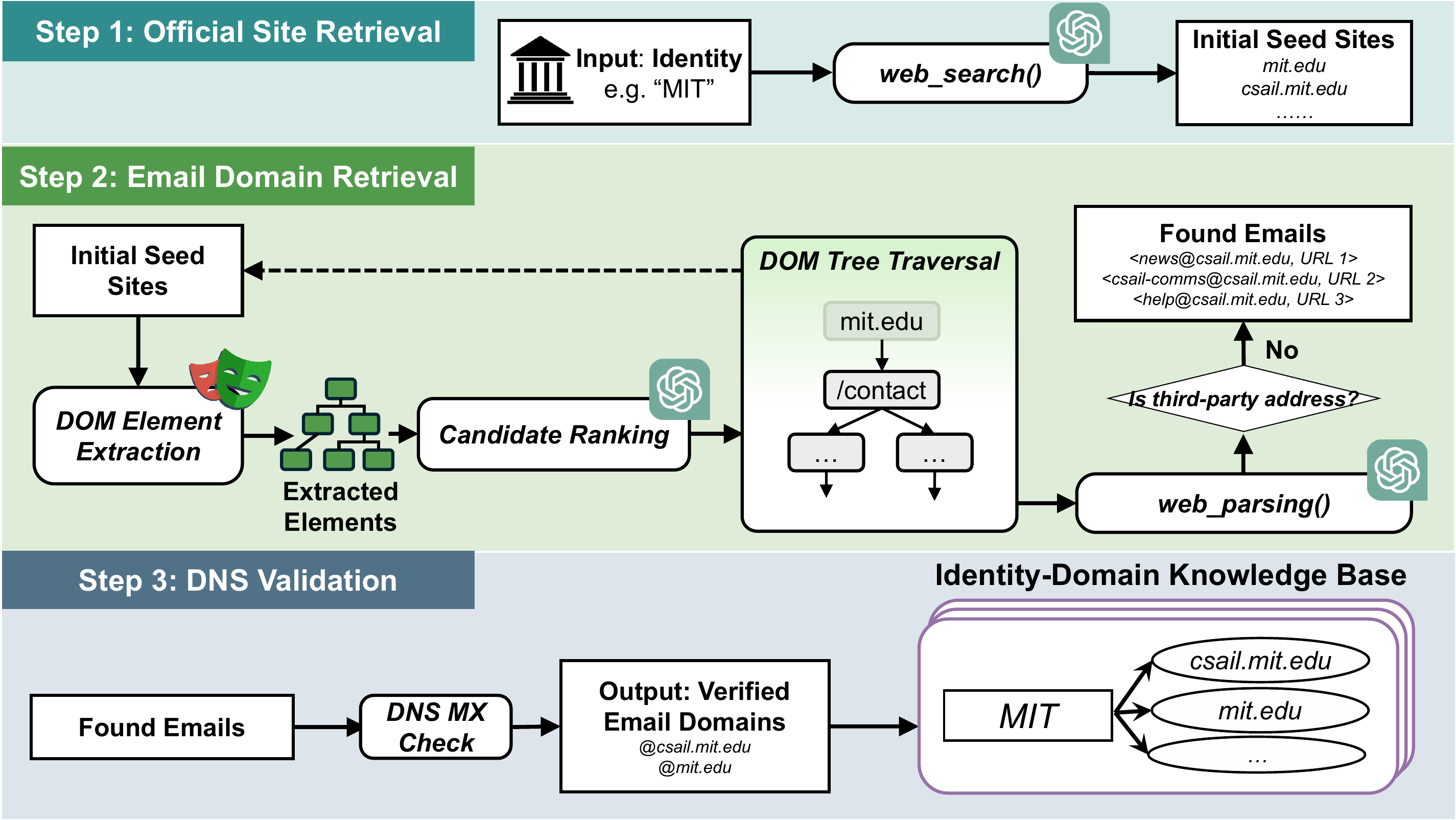}
    \caption{
    Knowledge Search Agent Implementation.
    }
    \label{fig:knowledge_agent}
\end{figure}

%\noindent\textbf{Knowledge Search Agent.}
We develop an email domain search agent to evolve the identity-domain knowledge base.
The input is an organization identity (e.g., ``Citibank''), and the output is a list of official contact email domains.
The agent workflow is shown in Figure~\ref{fig:knowledge_agent}.
Throughout the workflow, the agent is powered by OpenAI o4-mini~\cite{openai-o4mini}, which performs both the official-site retrieval in Step 1 and the page-traversal decisions in Step 2.

\begin{itemize}[leftmargin=*]
    \item \textbf{Step 1: Official Site Retrieval.}
    Given an identity $id$, we call an LLM with its built-in \textit{web\_search()} tool to retrieve a set of candidate official sites' domains $S_{id} = \{s_1, s_2, \dots\}$. 
    In cases where $id$ is ambiguous or corresponds to multiple entities  (e.g., ``Citibank'' is associated with \texttt{citi.com}, \texttt{citibank.com}, and \texttt{citigroup.com}), 
    we explicitly ask LLM to retain all candidate domains for downstream processing.
    
    \item \textbf{Step 2: Email Domain Retrieval.}
    For each candidate site $s \in S_{id}$, the agent performs page-level traversal to extract all listed email addresses $E_d = \{e_1, e_2, \dots\}$.
    We perform BFS from the homepage to extract interactive elements at each step and applying the LLM to choose which to expand.
    At every visited page, the agent invokes \textit{web\_parsing}(), which applies a regular expression to the page's HTML to extract email addresses.
    Traversal terminates upon reaching a predefined depth limit or when the LLM judges further expansion unproductive.

    \item \textbf{Step 3: Domain-Level DNS Validation.}
    We probe whether the retrieved emails $E_d$ have valid DNS MX records.
    This validation confirms that the recipient domain is operational.
    Operational email domains are added to the knowledge base.
    
\end{itemize}

\noindent\textbf{Initial Knowledge Base Construction.}
We curate an initial identity-domain knowledge base by running our knowledge search agent on the 5{,}600 phishing targets reported by DynaPhish~\cite{dynaphish} and KnowPhish~\cite{knowphish}.
For each input identity, the agent retrieves an average of 3.75 candidate official sites (Step 1),  
traverses each site to a maximum depth of 3 (Step 2), 
and applies DNS-level domain validation (Step 3), with an average end-to-end runtime of 7 minutes per identity.
This process yields a KB of 3,650 resolved identities mapped to 4,536 email domains.
Identities without sufficient supporting evidence are left unresolved rather than forcefully mapped.

\noindent\textbf{Knowledge Freshness.} 
Knowledge base entries can become outdated as organizations migrate or retire their email
domains. 
The construction process above can be rerun periodically offline, so entries can be re-validated and updated on a regular schedule.

% \subsubsection{User-Side Whitelisting.}\label{app:whitelisting}
% \R{2}{
% \rev{
% The agent-built knowledge base is global and domain-level: each entry maps an identity to its official email domains, shared by all users. 
% We complement it with a user-side layer at the sender-address level.
% Legitimate emails can carry a genuine identity-domain mismatch,
% e.g., an organizational announcement sent from a personal webmail
% account. 
% Such exceptions are inherently per-user and cannot be resolved by a global entry. 
% When \tool raises an alert, the user may confirm the sender as legitimate.
% A whitelist entry $\langle e, id \rangle$ binding the exact sender address $e$ to the claimed identity $id$ is then stored locally, and subsequent emails carrying the same pair are resolved as consistent, checked
% before the global knowledge base. The address-level granularity is
% necessary because a shared service domain sends on behalf of many
% organizations: a HotCRP-style platform delivers notifications for
% different conferences under one domain, so a domain-level entry
% would wrongly vouch for every tenant.
% This design follows the user-managed allowlists in production email systems, such as the Safe Senders list in Outlook~\cite{outlook-safe-senders},
% and never-send-to-spam filters in Gmail~\cite{gmail-spam-allowlist}.
% Note that if a whitelisted email address later claims a different identity, verification still applies. 
% We quantify the false-positive reduction and labeling effort in Section~\ref{exp:identity-domain-mismatched}.
% }
% }
\subsection{Instruction Recognition}\label{app:ins-recog}

\noindent\textbf{Definition.}
We define call-to-actions (CTAs) as phrases that prompt users to exit the current interface and perform a subsequent action, such as visiting an embedded link or QR code, replying to the email, or downloading attachments.

\noindent\textbf{Our Solution.}
To detect such actions, the solution is similar to claimed identity recognition: we employ a single NER model to label both claimed‐identity spans and action‐instruction spans in one pass.  
However, action instructions exhibit greater lexical variety than identity phrases. 
For instance, “\textit{click here}” may also appear as “\textit{follow this link}” or “\textit{visit this page}”.  
Raw email datasets may not include all these variations, and manually annotating additional data is labor-intensive.

To increase diversity, we adopt a structured way to augment the annotated CTAs.
Specifically, each training example has a 50\% chance of being paraphrased. 
The paraphrase is from sentence structures and persuasion strategies aspects.
We randomly select one of seven sentence structures: Imperative, Declarative-Permission, Fronted-Purpose, Conditional, Coordination, Relative-Clause, or Topic-Fronting~\cite{huddleston2002cambridge,biber1999longman}. 
Then we apply one of six persuasion strategies~\cite{cialdini1993psychology}: 
Reciprocity, Consistency, Social Proof, Authority, Liking, or Scarcity. 
For example, the original CTA is ``Click this link to verify your information''.
If ``Conditional'' and ``Scarcity'' are chosen, the paraphrased CTA could be:
``If you do not verify your information within the next 24 hours, you may lose access to your account, so please click this link now to secure it''.
The augmentation is conducted automatically using GPT-4o \cite{gpt4o} with the prompt shown in Supplementary Material D \cite{supplementary}.

% The resulting paraphrased email then serves as an updated training sample.
% This augmentation increases the diversity of words and sentence structures.
% Our adversarial experiment evaluation (Section \ref{exp:adv}) also shows that this data augmentation ensures robustness against paraphrasing attacks on unseen validation datasets. 
% The solution of instruction recognition is similar to sender identity recognition,
% where we can use the NER model to tag the tokens of call-to-action instruction.
% Therefore, we use one single NER model to output both claimed identity and instructions simultaneously.
% Compared to identity phrases, action phrases can be more diverse.
% For example, the simple instruction “click here” can be restated in multiple ways,
% such as “\textit{Follow this link}” or “\textit{Visit this page}”.
% Raw email datasets may not include all these variations, and manually annotating additional data is labor-intensive.

%\noindent \textbf{Instruction Augmentation.}

\section{\textit{SpearMail} Benchmark}\label{app:llm-benchmark}

% 围绕 k 个 properties，信息的细化程度，trust engineering property，人来打分的persuasiveness。维度有或者没有

To evaluate phishing detectors against the emerging threat of LLM-assisted spear-phishing, 
we construct \textit{SpearMail}, the largest profile-grounded spear-phishing benchmark generated against real-world researcher profiles. 
\textit{SpearMail} uniquely combines profile-grounded customization with high per-victim attack diversity (30 emails per profile).
Starting from \llmbenchmarkprofile\ publicly available researcher profiles, 
\textit{SpearMail} synthesizes \llmbenchmark\ personalized lures that impersonate senders from real, recognizable organizations 
and entice each target into performing attacker-specified actions.
Our generation pipeline is designed with three guiding principles: 
(1) \emph{customization}, 
(2) \emph{diversity} and 
(3) \emph{authenticity}.

\noindent\textbf{Profile Collection.}
We collect 681 researcher profiles from ORCID~\cite{orcid}, spanning PhD students, research fellows, and professors across 144 disciplines.
We choose researchers as the victim population for two reasons.
First, researcher profiles provide a controlled and verifiable source of public victim information: their affiliations, research interests, and professional activities are often explicitly documented in structured public records.
This allows us to objectively evaluate whether the generated emails are grounded in the victim's actual profile, rather than being generic phishing templates.
Second, although our benchmark uses researchers, the generation pipeline is not specific to academia.
It only requires public profile attributes, such as a person's role, interests, organization, and activities, which are also commonly available in other domains, such as corporate biographies, LinkedIn profiles, conference speaker pages, government directories, and professional webpages.
Therefore, researcher profiles serve as a measurable and ethically manageable proxy for studying profile-grounded spear-phishing, while the underlying attack mechanism generalizes to other populations.

% for the following reasons.
% Their academic activities (publications, affiliations, research interests) are systematically documented in structured public records, which 
% (i) enables our generation pipeline to ground emails in verifiable victim attributes and 
% (ii) allows us to objectively measure profile grounding against the original ORCID record as ground truth.
% But the generation pipeline itself is occupation-agnostic, and any structured public profile source would yield analogous emails for non-researcher victims.

\noindent\textbf{Email Generation.}
Directly prompting an LLM to produce a personalized phishing email from a raw profile tends to yield generic text that weakly references the victim.
We therefore decompose generation into three reasoning steps, $\textit{interest\_inference}$, $\textit{activity\_inference}$, and $\textit{email\_generation}$.
This forces the model to commit to a concrete recipient interest before writing. 
To suppress hallucinated organizations, we add a fact-checking step that verifies each generated organization exists and, when victim geolocation is available, that it is plausible for the victim's region. 
The full prompt templates are in Supplementary Material B \cite{supplementary}.

\begin{itemize}[leftmargin=*]
  \item \noindent\textbf{Step 1 (Interest Inference): }
    Given a parameter $m$, we infer $m$ potential interests from $p$, based on the prompt $\textit{interest\_inference()}$ in Supplementary Material Figure 9 \cite{supplementary}.
  \item \noindent\textbf{Step 2 (Activity Inference): }
    Given a parameter $n$, for each interest $i$, we infer $n$ relevant activities along with their corresponding organizations to attract the user, based on the prompt $\textit{activity\_inference()}$ in Supplementary Material Figure 9 \cite{supplementary}.
  \item \noindent\textbf{Step 3 (Email Generation): }
    Given the user $p$, a potential interest $i$, and a relevant activity $a$,
    we generate an email to invite $p$ with a pseudo-link in the name of $a$, based on the prompt $\textit{email\_generation()}$ in Supplementary Material Figure 9 \cite{supplementary}.
\end{itemize}

We set $m=6$ interests per profile and $n=5$ activities per
interest, yielding 30 candidate emails per victim and 20,430
generated emails in total.
The fact-checking step then discards 5,758 emails whose
impersonated organizations are hallucinated, i.e., non-existent, leaving the final \llmbenchmark\ emails impersonating 5{,}680 unique organizations.
Consider a user profile described as a ``PhD student with research interests in computational mathematics''.
Given this profile, one potential activity is an ``internship program at Google Research''.
The resulting spear-phishing email examples are present in Supplementary Material Figure 13 \cite{supplementary}.

\noindent\textbf{Ethics Consideration.}
\textit{SpearMail} is constructed entirely from publicly available ORCID profiles under ORCID's Public API Terms of Service, which permit non-commercial research use~\cite{orcid}. 
No generated email was delivered to any profiled researcher, no phishing infrastructure or live URLs were deployed, and our generation pipeline uses only benign prompts to commercial LLM APIs within the providers' standard terms of service. 
\textbf{To mitigate dual-use risk, the \textit{SpearMail} corpus will not be publicly disseminated, but will be made available to vetted researchers upon request under a data-use agreement.}
A detailed ethical statement is provided in Section~\ref{sec:ethics}.

% \begin{figure}
%     \centering
%     \fbox{
%     \includegraphics[width=0.9\linewidth]{figures/benchmark_eg.png}
%     }
%     \caption{An example in \dataset. The sensitive parts with recipient information and the recipient's previous publication are masked.}
%     \label{fig:benchmark_eg}
% \end{figure}

% \msgbox{Subject: Invitation to Google Internship Program}{

% Dear Dr. [\textit{Victim name}],

% We are pleased to inform you regarding an exciting opportunity for collaboration between your esteemed research group and Google Research.
% As we aim to strengthen the field of phishing detection technologies,
% we believe that your innovative insights and pioneering work,
% particularly in areas like [\textit{Victim's previous publication}],
% can provide invaluable direction for our upcoming internship program.

% For more information about the program structure and potential collaborative efforts, please don't hesitate to visit our detailed proposal here: [\textit{Suspicious link}]

% We are looking forward to possibly collaborating with you and igniting new learning opportunities for students eager to contribute to this critical field.

% Best regards,
% Google Research
% }

\noindent{\textbf{Benchmark Diversity.}}
We measure benchmark diversity by applying Latent Dirichlet Allocation (LDA) topic clustering~\cite{lda} to the \llmbenchmark\ emails, 
with the number of clusters selected via the elbow method on perplexity scores. 
This yields 500 distinct activity clusters and 150 interest clusters spanning multiple disciplines, including urban studies, healthcare, materials science, and energy.

\noindent{\textbf{Benchmark Customization.}}
To verify that \dataset constitutes \emph{spear}-phishing rather than generic phishing,
we measure the profile-grounding rate of each email against the recipient's original ORCID profile.
Among the \llmbenchmark\ emails, 
98\% reference the victim's research area
and 54\% mention at least one specific publication, 
confirming that \dataset\ emails are structurally grounded in victim profiles
rather than relying on generic phishing templates.

\section{Experiments}

%\subsection{Research Questions}
We carry out comprehensive experiments to answer the following research questions:
\begin{itemize}[leftmargin=*]
    \item \textbf{RQ1: Closed-World Experiment:}
    How effective can \tool detect phishing emails on different phishing benchmarks, compared to the state-of-the-art approaches?

    \item \textbf{RQ2: Open-World Experiment:}
    How does \tool perform on real-world emails in comparison to both academic baselines and industry-standard anti-spam filters?

    \item \textbf{RQ3: Adversarial Robustness Evaluation:}
    How resilient is \tool, a neural model based solution, against various adversarial attacks?

    \item \textbf{RQ4: Ablation Study:}
    How can each feature of \tool contribute to the performance of \tool?

    \item \textbf{RQ5: User Study:}
    Can \tool explanations help users better identify phishing emails?
\end{itemize}

\noindent{\textbf{Model Hyperparameter Setup}}
We train the NER model using the bert-large-uncased backbone released by Google \cite{bert}. 
The model is fine-tuned for 7 epochs with a learning rate of 2e-5 and a batch size of 8.
For the identity matching model, we directly use the same pre-training pipeline in \cite{characterbert-typo}. 
The CharacterBERT model has been pre-trained on English Wikipedia and OpenWebText \cite{liu2019roberta} and has been specifically designed to be resistant to typo-squatting attacks. 
We set the identity-matching threshold to 0.83 (Table~\ref{tab:threshold-selection}),
selected on the in-distribution training set
(Section~\ref{sec:dataset-train}) as the best
$\mathrm{F}_\beta$ trade-off, where we use $\beta=0.5$ to favor
precision and compute
$\mathrm{F}_\beta=\frac{(1+\beta^2)\,\mathrm{Precision}\cdot\mathrm{Recall}}
{\beta^2\,\mathrm{Precision}+\mathrm{Recall}}$.
All experiments were conducted on an Ubuntu 20.04 system using four NVIDIA RTX 4090 GPUs. 

\subsection{RQ1: Closed-World Experiment}\label{exp:closeworld}
\begin{table*}[h]
\centering
\caption{Performance on closed-world and real-world phishing benchmarks. 
Each cell shows the raw counts as well as FPR or Recall. 
Lower FPR is better; higher recall is better.
We highlight the top-3 solutions in green and the worst in red.}
\label{tab:closed-real-recall}
\resizebox{0.8\linewidth}{!}{%
\begin{tabular}{l|c|c|c|c}
\toprule
\textbf{Solution} 
& \makecell{\textbf{FP Counts}\\\textbf{(FPR on CSDMC)}} 
& \makecell{\textbf{TP Counts}\\\textbf{(Recall on Nazario \cite{nazario2005phishing})}} 
& \makecell{\textbf{TP Counts}\\\textbf{(Recall on PhishPot \cite{phishpot})}} 
& \makecell{\textbf{Runtime (s)}} \\
\midrule

D-Fence \cite{dfence} & 
\cellcolor{SoftRed}{2043 (97.15\%)} & 
\cellcolor{SoftGreen}{1697  (82.66\%)} & 
\cellcolor{SoftGreen}{788 (99.24\%)} & 
0.08 \\

HelpHed (Soft Voting) \cite{helphed} & 
\cellcolor{SoftGreen}{0 (0.00\%)} & 
1100  (53.58\%) & 
{213 (26.83\%)} & 
\textbf{0.03} \\

HelpHed (Stacking) \cite{helphed} & 
308 (14.65\%) & 
\cellcolor{SoftGreen}{1878  (91.48\%)} & 
685 (86.27\%) & 
\textbf{0.03} \\

ChatSpamDetector (GPT-4) \cite{chatspamdetector} & 
208 (9.89\%) & 
{752  (36.63\%)} & 
\cellcolor{SoftGreen}{794 (100.00\%)} & 
5.66 \\

SpamAssassin \cite{spamassassin} & 
\cellcolor{SoftGreen}{17 (0.81\%)} & 
\cellcolor{SoftRed}{51  (2.48\%)} & 
\cellcolor{SoftRed}{64 (8.06\%)} & 
1.32 \\

RSpamd \cite{rspamd} & 
83 (3.95\%) & 
1038  (50.56\%) & 
371 (46.73\%) & 
2.03 \\

\tool (Ours) & 
\cellcolor{SoftGreen}{25 (1.19\%)} & 
\cellcolor{SoftGreen}{1872  (91.18\%)} & 
\cellcolor{SoftGreen}{739 (93.07\%)} & 
0.04 \\

\midrule
\textbf{Total} & 
\textbf{2103} (100.00\%)& 
\textbf{2053} (100.00\%)& 
\textbf{794} (100.00\%)& 
-- \\

\bottomrule
\end{tabular}
}
\end{table*}

\begin{table*}[h]
\centering
\caption{Recall on LLM-generated phishing benchmarks. 
Each cell shows the raw counts as well as Recall. 
Higher recall is better.
We highlight the top-2 solutions in green and the worst in red.
*Note: E-PhishGen has 8{,}308 emails, but only 2{,}529 emails are imitating real organizations.}
\label{tab:llm-generated-recall}
\resizebox{\linewidth}{!}{%
\begin{tabular}{l|c|c|c|c}
\toprule
\textbf{Solution} 
& \makecell{\textbf{TP Counts}\\\textbf{(Recall on \dataset)}} 
& \makecell{\textbf{TP Counts}\\\textbf{(Recall on ContextualVector \cite{nahmias2024prompted})}} 
& \makecell{\textbf{TP Counts}\\\textbf{(Recall on E-PhishGen \cite{ephishgen})}} 
& \makecell{\textbf{TP Counts}\\\textbf{(Recall on SpearBot \cite{qi2025spearbot})}} \\
\midrule

D-Fence \cite{dfence} & 
\cellcolor{SoftRed}{0 (0.00\%)} & 
\cellcolor{SoftRed}{0 (0.00\%)} & 
\cellcolor{SoftRed}{0 (0.00\%)} & 
\cellcolor{SoftRed}{0 (0.00\%)} \\

HelpHed (Soft Voting) \cite{helphed} & 
\cellcolor{SoftRed}{0 (0.00\%)} & 
52 (15.57\%) & 
1155 (45.67\%) & 
1 (0.10\%) \\

HelpHed (Stacking) \cite{helphed} & 
\cellcolor{SoftGreen}{14132 (96.32\%)} & 
254 (76.05\%) & 
1450 (57.33\%) & 
\cellcolor{SoftGreen}{416 (41.60\%)} \\

ChatSpamDetector (GPT-4) \cite{chatspamdetector} & 
11150 (76.00\%) & 
\cellcolor{SoftGreen}{334 (100.00\%)} & 
\cellcolor{SoftGreen}{2259 (89.32\%)} & 
142 (14.20\%) \\

SpamAssassin \cite{spamassassin} & 
22 (0.15\%) & 
\cellcolor{SoftRed}{0 (0.00\%)} & 
8 (0.32\%) & 
\cellcolor{SoftRed}{0 (0.00\%)} \\

RSpamd \cite{rspamd} & 
42 (0.29\%) & 
\cellcolor{SoftRed}{0 (0.00\%)} & 
\cellcolor{SoftRed}{0 (0.00\%)} & 
1 (0.10\%) \\

\tool (Ours) & 
\cellcolor{SoftGreen}{12678 (86.41\%)} & 
\cellcolor{SoftGreen}{301 (90.12\%)} & 
\cellcolor{SoftGreen}{2113 (83.55\%)} & 
\cellcolor{SoftGreen}{841 (84.10\%)} \\

\midrule
\textbf{Total} & 
-- / \textbf{14672} & 
-- / \textbf{334} & 
-- / \textbf{2529*} & 
-- / \textbf{1000} \\
\bottomrule
\end{tabular}
}
\end{table*}

% \begin{table*}[h]
%     \caption{Experimental results on closed-world datasets. 
%     We calculate the false positive rate on the benign email dataset, i.e., CSDMC; recall (i.e., false negative) on the phishing email datasets, i.e., Nazario, PhishPot, and SpearMail; and median runtime in seconds.}
%     \label{tab:close-world}
%     \centering
%     \resizebox{\linewidth}{!}{\begin{tabular}{l|c|c|c|c|c}
%     \toprule
%      \textbf{Solutions} &  \textbf{FPR on CSDMC} & \textbf{Recall on Nazario} & \textbf{Recall on PhishPot} & \textbf{Recall on SpearMail} & \textbf{Runtime} \\
%      \midrule
%      \midrule
%      D-Fence \cite{dfence} & 97.15\%	& 82.66\%	& \textbf{99.21\%} &  0.00\%	& 0.08 \\

%      HelpHed (Soft Voting) \cite{helphed} & 	2.00\% & 	53.58\%	& 26.32\% & 0.00\%	& \textbf{0.03} \\

%     HelpHed (Stacking) \cite{helphed} & 	14.65\%	& 50.75\% & 85.92\% & \textbf{99.69\%}	& \textbf{0.03} \\

%     ChatSpamDetector (GPT4)	\cite{chatspamdetector} & 14.88\% & 	\textbf{98.99\%}	& \textbf{99.75\%} & 58.95\% & 	5.66 \\

%     SpamAssassin \cite{spamassassin} & \textbf{0.81\%} & 2.68\% & 7.81\% & 0.15\% & 1.32 \\
%     RSpamd \cite{rspamd} & 3.95\% & 49.98\% & 46.22\% & 0.29\% & 2.03 \\

%      \tool & \textbf{1.19\%} &	\textbf{91.18\%}	& 86.02\% & \textbf{99.02\%} & \textbf{0.04} \\

%     \bottomrule
%     \end{tabular}
%     }

% \end{table*} 

\subsubsection{\textbf{Baselines}}
We select baselines by three criteria: coverage of the major design paradigms for phishing detection, representativeness measured by adoption in follow-up work and citations, and availability of a runnable implementation for faithful reproduction.
This yields three groups, described in detail in Supplementary Material Section F \cite{supplementary}.
\textbf{Feature-engineering detectors} aggregate handcrafted content features, for which we use HelpHed~\cite{helphed} and D-Fence~\cite{dfence} under the configurations recommended in the original papers, including both HelpHed ensembling variants.
The \textbf{LLM-based detector} ChatSpamDetector~\cite{chatspamdetector} performs prompt-driven reasoning over cues such as impersonation and suspicious actions or links, and we use GPT-4 following the original setup.
The \textbf{production anti-spam filters} RSpamd~\cite{rspamd} and SpamAssassin~\cite{spamassassin} are widely deployed rule- and reputation-based systems that approximate real-world deployments, run under their default configurations.

\subsubsection{\textbf{Setup}}\label{sec:dataset}
\noindent\textbf{Training and In-Distribution Test.}\label{sec:dataset-train}
Following D-Fence and HelpHed~\cite{dfence, helphed}, we train all learning-based methods on the same corpora, 4,558 phishing emails from the Nazario 2005 corpus~\cite{nazario2005phishing} and 10,000 benign emails subsampled from Enron~\cite{enron}.
For NER training we manually annotate a random subset of 2,086 emails from these corpora with claimed identities and CTAs, since labeling all is prohibitive, and split it into 1,701 for training and 385 for testing.
Two annotators are involved, the first performing the initial labeling and the second independently reviewing and correcting it in Label Studio~\cite{labelstudio}, with 83\% agreement and disagreements resolved by the reviewer.
All methods, including \tool, are trained on this conventional corpus and never on \dataset, ensuring a fair comparison.

\noindent\textbf{Conventional Benchmarks (Out-of-Distribution Test).}\label{sec:dataset-conventional}
To evaluate generalization to a different distribution, we use a separate test set from newer and independent sources, 2,584 phishing emails from the Nazario corpus (2015--2023) and 2,949 benign emails from CSDMC-Ham~\cite{csdmc}, yielding 2,053 phishing and 2,103 benign emails after deduplication.
We also include results on the benign splits of TREC-05/06/07~\cite{trecspam}, CEAS-08~\cite{ceas2008}, Enron~\cite{enron}, SpamAssassin~\cite{spamassassincorpus}, and Ling-Spam~\cite{lingspam}.

\noindent\textbf{Real-World and LLM-Generated Phishing.}
We further evaluate on \textit{PhishPot}~\cite{phishpot}, an open repository of real-world phishing emails collected between Sep 2023 and Nov 2024, from which 4,300 collected emails yield 794 unique phishing emails after deduplication, and on four LLM-generated benchmarks, \textit{Prompted Contextual Vector}~\cite{nahmias2024prompted}, \textit{E-PhishGen}~\cite{ephishgen}, \textit{SpearBot}~\cite{qi2025spearbot}, and \textit{SpearMail} (Section~\ref{app:llm-benchmark}).
These four contain profile-grounded spear phishing, whereas Nazario and PhishPot represent generic phishing with identity claims, so together they cover both classes in our scope (Section~\ref{sec:threat-model}).

\noindent\textbf{Metrics.}
We report the False Positive Rate (FPR) on the CSDMC dataset, recall on the phishing datasets, and the median runtime as a measure of operational cost.

\subsubsection{\textbf{Results of Phishing Detection}}
\autoref{tab:closed-real-recall} and \autoref{tab:llm-generated-recall} show that \tool achieves a consistent advantage over the baselines in balancing false positives, false negatives, and runtime overhead.
The reported median runtime of 0.05 seconds covers per-email identity recognition and knowledge base matching.
It excludes knowledge base expansion (Section~\ref{app:knowledge-construction}), which is triggered only on a cold miss, i.e., the first encounter of an unresolved identity, runs offline and asynchronously, and is a one-time cost per identity of about 7 minutes amortized across all subsequent emails claiming that identity.
Feature-engineering solutions struggle to balance false positives and negatives, largely due to distribution shift.
ChatSpamDetector performs well on PhishPot but is over-aggressive on benign emails, and it retains non-negligible false negatives on LLM-generated phishing because it makes ungrounded decisions on whether an email is \textit{impersonated}.
RSpamd and SpamAssassin rarely produce false alerts but miss the majority of phishing emails, even on conventional datasets.

\begin{figure*}[t]
    \centering
    \begin{subfigure}{0.3\textwidth}
        \fbox{\includegraphics[width=\textwidth, height=1.7cm]{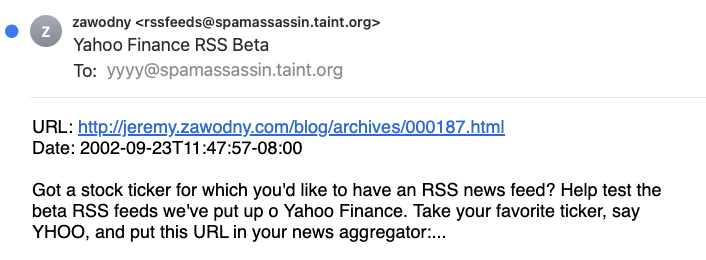}}
    \subcaption{FP example on CSDMC dataset}\label{fig:csdmc-fp}
    \end{subfigure}
    \hfill
    \begin{subfigure}{0.3\textwidth}
        \fbox{\includegraphics[width=\textwidth, height=1.7cm]{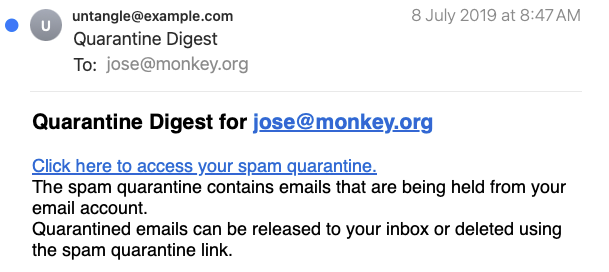}}
        \subcaption{FN example on Nazario dataset}\label{fig:nazario-fn}
    \end{subfigure}
        \hfill
    \begin{subfigure}{0.3\textwidth}
        \fbox{\includegraphics[width=\textwidth, height=1.7cm]{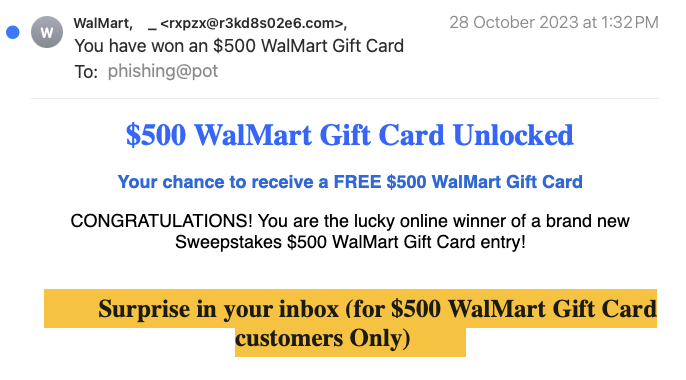}}
        \subcaption{FN example on PhishPot dataset}\label{fig:phishpot-fn}
    \end{subfigure}
    \caption{Failure examples of \tool}
\end{figure*}
%\subsubsection{Qualitative Studies}

\noindent\textbf{Why do D-Fence and HelpHed overfit?}
\label{sec:overfit-discussion}
Both D-Fence and HelpHed can overfit because their decisions are dominated by a few header- or format-level features that are not stable across deployments. 
We visualize the top-3 important features of them in Supplementary Material Figure 11 \cite{supplementary}.
We observe that D-Fence heavily relies on the count of ``Received'' headers.
D-Fence operates under the assumption that benign emails typically traverse multiple layers of email servers.
However, legitimate emails can vary widely in their routing paths,
and malicious actors could easily manipulate the ``Received'' headers to mimic benign patterns.
HelpHed can be biased toward Content-Transfer-Encoding.
While encoding type can offer some insights into the nature of the email content, it is not a definitive indicator of phishing.

\noindent\textbf{Why does ChatSpamDetector have false positives and negatives?}\label{sec:hallucination-discussion}
ChatSpamDetector makes mistakes when it is forced to make \textit{ungrounded} decisions on whether an email is impersonated.
Specifically, it detects phishing emails using a \textit{single} information source.
%is likely to produce hallucinated responses.
For instance, when the phishing email is imitating the ``Annual Conference on Human-Robot Interaction (HRI 2025)'',
if a sender address is registered as \textit{ieeehri@humanrobot.com},
GPT might mistakenly recognize it as consistent with the identity.
In contrast, the official address should be from \textit{humanrobotinteraction.org},
which can be well captured by the reference-based design of \tool.
More examples of false negatives and false positives can be found on our anonymous website \cite{anonymoussupeg}.

%This scenario underscores the hallucination issues associated with the direct use of large language models (LLMs) for tasks requiring precise and up-to-date information.
%To address these limitations, it is crucial to incorporate an explicit knowledge base as a reference for verifying sender addresses.

\noindent\textbf{Why do RSpamd and SpamAssassin miss phishing emails?}\label{sec:rspamd-discussion}
We examine the rules that most often trigger each solution.
SpamAssassin fires mainly on ``MIME HTML ONLY'', ``TO MALFORMED'', and ``DKIM SIGNED'', while RSpamd fires on ``DATE IN PAST'', ``RDNS NONE'', and ``MANY INVISIBLE PARTS''.
These rules key on transport and formatting artifacts rather than the email's claims, so they are interpretable but trivially circumvented by an attacker who sends well-formed, DKIM-signed mail.

\noindent\textbf{\tool False Negatives.}
\noindent\textbf{(1) Missing or ambiguous sender identities.}
As illustrated in \autoref{fig:nazario-fn}, some phishing emails avoid claiming a clear organizational identity.
In such cases, \tool does not recognize any claimed identity, and therefore cannot perform the downstream inconsistency check.
This reflects an ASR (Attack-Success-Rate) and BR (Bypass-Rate) trade-off: vague identities may evade detection yet reduce user trust~\cite{khadka2023survey}.
As a mitigation, when \tool fails to extract a confident sender identity, we plan to extend identity inference from richer contextual cues in future work.
Moreover, instead of treating such cases as benign, we can explicitly flag ``missing or highly ambiguous identity'' as a soft warning signal to downstream filters or user interfaces.
\noindent\textbf{(2) Non-salient call-to-action phrases.}
As shown in \autoref{fig:phishpot-fn}, some phishing emails use CTAs such as ``Surprise in your inbox'' that stand out visually but lack an explicit imperative verb, and thus are not recognized as clear instructions.
To mitigate this, we can incorporate visual saliency detection~\cite{itti2002model,bruce2009saliency} into our pipeline, combined with cognitive and advertising theories on how visually prominent elements drive behavior~\cite{pieters1999visual,cialdini1993psychology}.
Such extensions would help \tool better capture indirect yet visually highlighted CTAs.

% There are two primary reasons for false negatives:
% (i) \textit{Spoofed sender address}:
% The sender's address is consistent with the claimed identity, but the address is spoofed.
% This issue can be effectively mitigated by implementing SPF checks.

% (ii) \textit{Ambiguous identity}:
% The emails exhibit missing or ambiguous sender identities.
% As illustrated in \autoref{fig:nazario-fn}, 
% \tool does not recognize any claimed identity from the email,
%humans can infer that the email impersonates an internal role, such as a mail administrator.
%However, since explicit indicative phrases like ``mailbox admin'' are absent, 
% thus \tool cannot follow up with the inconsistency check between the detected identity and the domain.
%fails to detect any sender identity.
% This highlights a potential dilemma for attackers between email clarity and the success rate of their attacks.
% By deliberately adopting a vague identity, attackers may evade the detection, 
% but may lead to less effective phishing attempts.

% This problem can be mitigated by augmenting more call-to-action training samples.

% Some emails only include

\noindent\textbf{\tool False Positives}
Upon investigation, we find that part of \tool's false alarms on the CSDMC benign dataset stem from label noise: the dataset itself contains some suspicious
emails. 
Figure~\ref{fig:csdmc-fp} is an example \tool\ flags as Yahoo Finance
phishing, promoting a beta Yahoo Finance RSS feed with a test URL
but sent from \texttt{rssfeeds@spamassassin.taint.org}, which does
not belong to the Yahoo domain. 
We provide a systematic analysis of \tool's false positives on a large public benign corpus in Section~\ref{sec:fp-stress}.

\subsubsection{\textbf{Understanding False Positives on Benign Emails}}
\label{sec:fp-stress}
\begin{figure*}[tp]
\centering

% \vspace{6pt}
% ---- Row 2: genuine identity-domain inconsistency ----
\begin{subfigure}[t]{0.32\linewidth}
  \centering
  \fbox{\includegraphics[height=1.7cm, width=\linewidth]{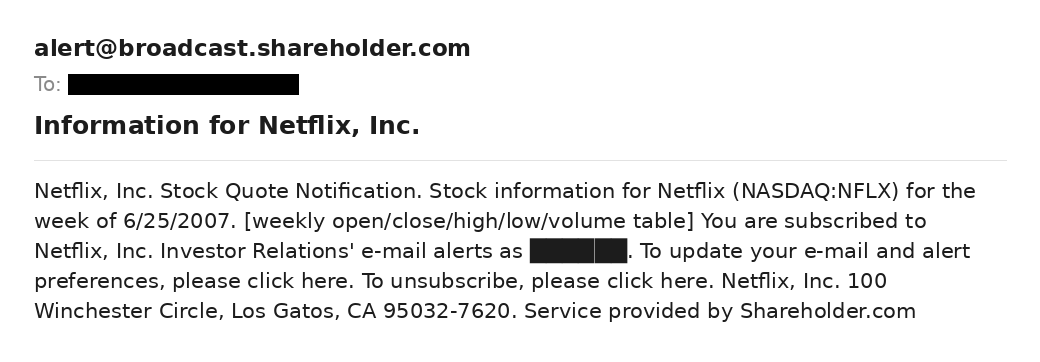}}
  \caption{A Netflix investor-relations alert delivered by the
  third-party service \texttt{shareholder.com}.}
  \label{fig:fp-genuine-netflix}
\end{subfigure}\hfill
\begin{subfigure}[t]{0.32\linewidth}
  \centering
  \fbox{\includegraphics[height=1.7cm, width=\linewidth]{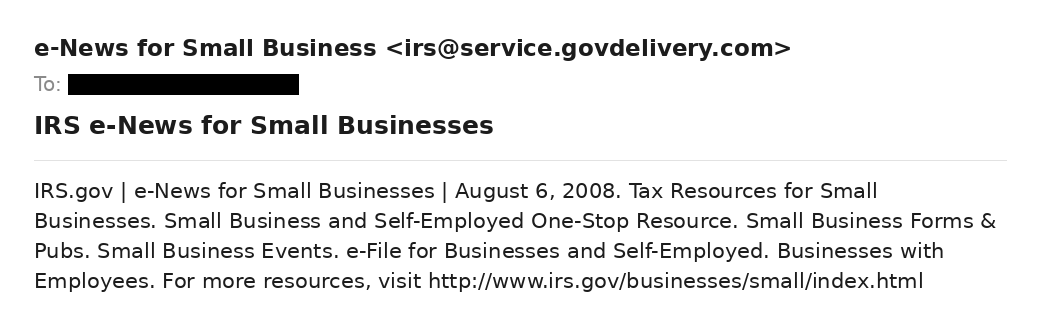}}
  \caption{A genuine IRS newsletter distributed through the
  third-party service \texttt{govdelivery.com}, used by many
  government agencies, rather than an \texttt{irs.gov} domain.}
  \label{fig:fp-genuine-irs}
\end{subfigure}\hfill
\begin{subfigure}[t]{0.32\linewidth}
  \centering
  \fbox{\includegraphics[height=1.7cm, width=\linewidth]{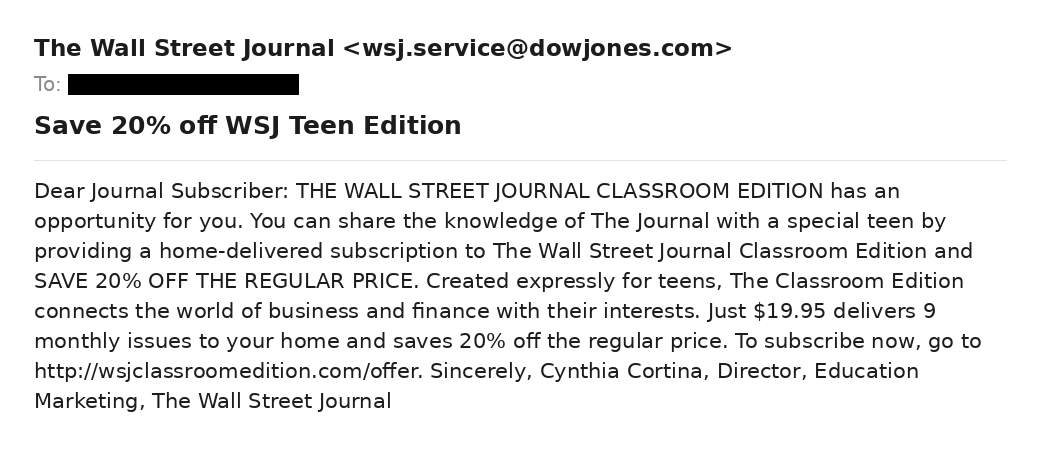}}
  \caption{A genuine Wall Street Journal newsletter sent from its
  parent company's domain \texttt{dowjones.com} rather than an
  official WSJ domain.}
  \label{fig:fp-genuine-wsj}
\end{subfigure}
\caption{
False-positive cases from \tool\ on the public benign corpus: 
they exhibit genuine identity-domain inconsistencies, where a legitimate brand delivers mail through a third-party service or parent-company domain rather than its own official domain. Recipient fields are redacted for
privacy.
}
\label{fig:fp-analysis}
\end{figure*}

To understand when \tool\ raises false alarms, we run it on the benign portion of the seven-dataset email collection~\cite{sevenphishingdatasets}, which aggregates 108{,}689 benign emails from TREC-05/06/07~\cite{trecspam}, CEAS-08~\cite{ceas2008}, Enron~\cite{enron}, SpamAssassin~\cite{spamassassincorpus}, and Ling-Spam~\cite{lingspam}.
Of these, 90{,}155 carry a sender address and form our evaluation set, on which \tool\ flags 250 emails, a false positive rate of \textbf{$0.28\%$}.
These emails exhibit \textit{genuine identity-domain inconsistencies}, typically because the email is delivered through a third-party Email Service Platform, sent from a sibling or parent domain of the claimed organization, or sent on behalf of an organization from a personal address (Figure~\ref{fig:fp-analysis}).
This reveals a limitation of \tool, since our knowledge base is \textit{domain-level} whereas a specific sender address is sometimes legitimately authorized to represent an identity, e.g., \texttt{\seqsplit{wsj.service@dowjones.com}} sends newsletters on behalf of the Wall Street Journal.
A natural remedy is an address-level whitelist layered on the domain-level knowledge base, following the user-managed allowlists in production email systems~\cite{outlook-safe-senders, gmail-spam-allowlist}, binding an exact sender address $e$ to a claimed identity $id$ and exempting only that pair $\langle e, id\rangle$.
Such a whitelist is cheap in practice, as the false positives are concentrated on a few high-volume senders.
Whitelisting the \textbf{Top-10} most frequent senders already eliminates $216$ of the $250$ false positives ($86.4\%$) while introducing no false negatives, since an exemption is scoped to the exact pair and the same address claiming a \textit{different} identity is still flagged.
We leave a full design to future work, where the open questions concern how entries are bootstrapped and revoked when a delegated address changes ownership or is compromised, and how the whitelist interacts with authentication signals such as SPF and DKIM alignment so that an exemption is never granted to a spoofable address.

\subsubsection{\textbf{Does \tool Overfit the \dataset Template?}}
\label{sec:template-invariance}
Since \dataset is produced by a fixed LLM generation pipeline, its emails may share a common template, and a detector could appear effective by recognizing the template rather than the underlying phishing signal.
We examine this from both directions.
On the phishing side, we sample 500 \dataset emails and instruct GPT-4o to perturb their discourse structure, reordering the functional segments (greeting, pretext, call-to-action, sign-off) and relocating the embedded link, while preserving the claimed identity, the recipient-specific details, the URL string, and the phishing intent.
On the benign side, we sample 500 CSDMC-Ham emails and rewrite them into \dataset's presentation style while strictly preserving their communicative intent.
Both prompts are in Supplementary Material Section C \cite{supplementary}.
A template-dependent detector should produce false negatives and false positives respectively.
Table~\ref{tab:template-invariance} shows that \tool retains a recall of 81.0\% under discourse perturbation and reports 0.0\% false positives on the rewritten benign emails, indicating that its decisions rest on the counterfactual identity-domain inconsistency rather than on the generation template, whereas the baselines either fail to detect the perturbed phishing emails or become unstable on the template-styled benign ones.

\begin{table}[t]
\caption{
Template-invariance evaluation. Recall on 500
discourse-perturbed SpearMail emails and FPR on 500 benign CSDMC-Ham
emails rewritten into SpearMail's surface style.
}
\label{tab:template-invariance}
\centering
\resizebox{\linewidth}{!}{%
\begin{tabular}{l|c|c}
\toprule
\textbf{Solution} & \makecell{\textbf{Recall}\\\textbf{(500 Perturbed \dataset)}} & \makecell{\textbf{FPR}\\\textbf{(500 Rewritten CSDMC-Ham)}} \\
\midrule
D-Fence                  & 0.00\%  & 0.00\% \\
HelpHed (Soft Voting)    & 0.00\%  & 1.40\% \\
HelpHed (Stacking)       & 1.60\%  & 28.20\% \\
ChatSpamDetector (GPT-4) & 47.60\% & 0.60\% \\
SpamAssassin             & 0.00\%  & 0.00\% \\
RSpamd                   & 0.00\%  & 2.20\% \\
\tool (Ours)   & \textbf{81.00\%} & \textbf{0.00\%} \\
\bottomrule
\end{tabular}
}
\end{table}

\subsubsection{\textbf{Results of Knowledge Evolution}}
To evaluate the agent's output quality, we randomly sample 100 identities and manually collect their official contact emails from organization contact pages, LinkedIn, and verified social media as ground truth.
The agent achieves 97\% precision and 93\% recall on this subset, confirming that the construction strategy yields suitable performance.
False positives mainly arise from anti-crawling obfuscation, where official pages hide contact emails in non-standard forms or image-rendered addresses that can be mistakenly converted into valid-looking domains.
False negatives mainly arise from the limited navigation capability of the LLM-based search agent, as some official addresses sit several clicks from the homepage, under dynamically rendered pages, inside PDFs, or across department-level subpaths.
Both can be mitigated in future work with more engineering effort.
\subsection{RQ2: Open-World Experiment}\label{exp:openworld}

In this study, we investigate the performance of \tool and baselines for phishing emails occurring in the wild.
We present the dataset summary we used in this experiment in Table~\ref{tab:openworld-dataset}.

\begin{table}[h]
    \caption{Open-world datasets summary.}
    \label{tab:openworld-dataset}
    \centering
    \resizebox{\linewidth}{!}{
    \begin{tabular}{c|c|c|c}
    \toprule
    Datasets & Total & \makecell{\# Wild \\ Phishing Emails} & \makecell{\# Simulated \\ Phishing Emails} \\
    \midrule
    Researcher Email Dataset & 9,369 & 6 (reported) & 87  \\
    University Spam Feeds & 2,850 & 1,868 & 0 \\
    Honeypot Phishing & 70 & 70 & 0 \\
    \bottomrule
    \end{tabular}
    }
\end{table} 
\begin{table*}[h]
\centering
\caption{
Experimental results on open-world datasets. }
\label{tab:open-world}
\resizebox{0.9\linewidth}{!}{%
\begin{tabular}{l|c|c|c|c|c|c}
\toprule
\textbf{Solution} 
& \makecell{\textbf{Researcher Email}\\\textbf{Precision}} 
& \makecell{\textbf{Researcher Email}\\\textbf{Recall (Simulated)}} 
& \makecell{\textbf{University Spam Feed Dataset}\\\textbf{Recall (Wild)}} 
& \makecell{\textbf{Honeypot Phishing Dataset}\\\textbf{Recall (Wild)}} 
& \makecell{\textbf{Runtime}\\\textbf{(s)}} \\
\midrule

D-Fence \cite{dfence} & 
1.39\% & 
\textbf{100.00\%} & 
79.49\% & 
\textbf{100.00\%} & 
0.11 \\

HelpHed (Soft Voting) \cite{helphed} & 
2.57\% & 
17.24\% & 
44.22\% & 
32.86\% & 
0.08 \\

HelpHed (Stacking) \cite{helphed} & 
0.42\% & 
20.69\% & 
62.52\% & 
90.00\% & 
0.07 \\

ChatSpamDetector (GPT-4) \cite{chatspamdetector} & 
6.85\% & 
93.10\% & 
\textbf{98.07\%} & 
95.71\% & 
3.75 \\

University Anti-Spam Filter & 
-- & 
3.45\% & 
-- & 
-- & 
-- \\

\tool & 
\textbf{90.29\%} & 
\textbf{100.00\%} & 
80.51\% & 
87.14\% & 
0.05 \\

\bottomrule
\end{tabular}
}
\end{table*}

\noindent\textbf{University Spam Feed dataset.}
The University provides access to a subscription-based feed available to researchers, which logs emails received within the university organization that are flagged by its anti-spam filter (RSpamd \cite{rspamd}). 
This dataset spans 2024-11 to 2026-04 and comprises 2,850 spam
emails, of which
we label 1,868 as phishing. The labeling follows our
threat-model definition (Section~\ref{sec:threat-model}), i.e., an
email is phishing if it both impersonates an identity and carries
a call-to-action, and is applied by the authors via manual
inspection, independent of \tool's output.
This openly accessible dataset is included as part of our
evaluation.

\noindent\textbf{Honeypot Phishing dataset.}
In addition, we registered a honeypot email account, which was actively distributed by submitting it to phishing websites. 
Each day, the email address was submitted to 100 newly identified phishing sites listed on OpenPhish \cite{openphish} using an automated Selenium-based form filler \cite{dynaphish}. 
This honeypotting activity was conducted over the course of one year, attracting 70 phishing emails. 
This activity involved only the passive collection of unsolicited phishing emails.
\textbf{No personally identifiable information was collected}, and all procedures adhered to applicable ethical and institutional guidelines.

\noindent\textbf{Real-World Researcher Email Evaluation.}
To assess deployment feasibility, we conducted a small-scale on-device field evaluation 
with three consented volunteers at the same institution (IRB No.~20250859I).
\tool ran locally on participants' machines; no raw email content or identifiers 
were collected or transmitted.
Participants provided feedback only for emails flagged as suspicious by \tool or a baseline.
To supplement the rarity of naturally occurring phishing emails,
each participant additionally received
30 synthetic spear-phishing emails generated via the
same procedure as SpearMail.
3 of the 90 were dropped for referencing non-existent organizations, yielding 87 in total.
None of the three participants was involved in the design of \tool.
Participants consented in advance to receive simulated phishing emails during the study period, but were not told their timing, sender, or content.
None reported acting on any synthetic email (e.g., clicking its link or replying).
For internal-identity matching (Section~\ref{app:identity-matching}), the recipient domain is taken directly as the volunteer's own mailbox domain, without additional inferences.

\subsubsection{\textbf{Results}}
We evaluate the performance of each solution using precision and recall.
As shown in \autoref{tab:open-world}, we observe that D-Fence, ChatSpamDetector, and HelpHed tend to report false alarms.
On the researcher's email dataset, \tool is less likely to report false alarms.
Our approach achieves recalls of 80.5\% and 87.1\% on the spam feed and honeypot datasets, respectively.
More wild phishing examples are shown in Appendix~\ref{sec:vlm-phishing}.

\subsection{RQ3: Adversarial Robustness}\label{exp:adv}

\subsubsection{\textbf{Attack Setup}}
Once deployed, \tool's model weights are confidential, thus, 
we evaluate the robustness of \tool under practical, black-box text perturbations widely used in prior work \cite{bae, deepwordbug, textflint-concatsent, textfooler, holgers2006cutting, modular}. 
We consider three attack surfaces: 
(1) \textbf{email processing} via text salting/invisible-character injection \cite{text-salt}, 
(2) \textbf{identity and CTA extraction (the NER model)} via token and character perturbations and paraphrasing (BAE \cite{bae}, DeepWordBug \cite{deepwordbug}, GPT paraphrasing, ConcatSent \cite{textflint-concatsent}, TextFooler \cite{textfooler}) and 
(3) \textbf{identity matching} via typo-level perturbations on brand names (DeepWordBug \cite{deepwordbug}). 
Surfaces (1–2) are evaluated on the 385 test emails in Section~\ref{sec:dataset-train}, and surface 
(3) on the 4,536 knowledge-base brand variants (Section~\ref{app:knowledge-construction}).

\noindent\textbf{Attack Surface 1: Email Processing.}
We apply the text salting attack \cite{text-salt}, where emails are manipulated by inserting invisible characters at random positions. 
After injection, we assess whether the rendered text contains these “salted” characters.

\noindent\textbf{Attack Surface 2: Identity NER.} 
For attacks on the named entity recognition (NER) model, we differentiate between identity and action classes, as action phrases tend to be longer and semantically richer.
For the \textit{identity class}, we apply:
\begin{itemize}[leftmargin=*]
\item \textbf{BAE} \cite{bae}: Masks the token immediately preceding the target entity and uses a pre-trained BERT model to generate top-k candidate insertions.
\item \textbf{DeepWordBug} \cite{deepwordbug}: Introduces character-level perturbations by replacing, deleting, switching, or repeating characters within an entity.
\end{itemize}
For the \textit{action class}, we apply:
\begin{itemize}[leftmargin=*]
\item \textbf{GPT Paraphrasing}: Rewrites the action phrase using a paraphrasing model.
\item \textbf{ConcatSent} \cite{textflint-concatsent}: Merges phrases with preceding sentences to disrupt original sentence boundaries.
\item \textbf{TextFooler} \cite{textfooler}: Replaces key verbs with contextually appropriate synonyms.
\end{itemize}
We evaluate the entity recognition rate to determine whether the model can still accurately identify and classify entities under adversarial perturbations.

\noindent\textbf{Attack Surface 3: Identity matching.} 
To assess robustness against attacks on the identity matching model, we evaluate the matching accuracy between original and perturbed brand names using the \textit{DeepWordBug} \cite{deepwordbug} attack.

%, as detailed in Section \ref{app:knowledge-construction}.

\subsubsection{\textbf{Results}}

\begin{table}[t]
\centering
\caption{Adversarial attacks on the three components of \tool.
Panel (a) reports attack success rate, where $0\%$ means the salting characters still appear in the rendered text seen by \tool.
Panels (b) and (c) report recognition and matching rates as percentages.}
\label{tab:adv_results}
\footnotesize
\setlength{\tabcolsep}{4pt}
\begin{tabular}{l|cc}
\toprule
\multicolumn{3}{l}{\textbf{(a) Email processing.}} \\
\textbf{Attack} & \multicolumn{2}{c}{\textbf{Attack Success Rate}} \\
\midrule
Zero-width characters \cite{text-salt}      & \multicolumn{2}{c}{0\%} \\
Hidden CSS elements \cite{text-salt}        & \multicolumn{2}{c}{0\%} \\
White text on white background \cite{text-salt} & \multicolumn{2}{c}{0\%} \\
\midrule\midrule
\multicolumn{3}{l}{\textbf{(b) NER model.}} \\
\textbf{Attack (target field)} & \textbf{Clean} & \textbf{After Attack} \\
\midrule
BAE \cite{bae} (identity)                        & 89\% & 87\% {\small($\downarrow$2.25\%)} \\
DeepWordBug \cite{deepwordbug}--Delete (identity)  & 91\% & 92\% {\small($\uparrow$1.10\%)} \\
DeepWordBug \cite{deepwordbug}--Replace (identity) & 94\% & 94\% {\small(--)} \\
DeepWordBug \cite{deepwordbug}--Switch (identity)  & 90\% & 91\% {\small($\uparrow$1.11\%)} \\
DeepWordBug \cite{deepwordbug}--Repeat (identity)  & 92\% & 92\% {\small(--)} \\
GPT Paraphrase (action)                          & 90\% & 88\% {\small($\downarrow$2.22\%)} \\
ConcatSent \cite{textflint-concatsent} (action)  & 90\% & 89\% {\small($\downarrow$1.11\%)} \\
TextFooler \cite{textfooler} (action)            & 89\% & 89\% {\small(--)} \\
\midrule\midrule
\multicolumn{3}{l}{\textbf{(c) Identity matching model.}} \\
\textbf{Attack} & \textbf{BERT} & \textbf{CharBERT} \\
\midrule
No Attack                                 & 100\% & 100\% \\
DeepWordBug \cite{deepwordbug}--Delete    & 22\% {\small($\downarrow$78\%)} & 73\% {\small($\downarrow$27\%)} \\
DeepWordBug \cite{deepwordbug}--Replace   & 22\% {\small($\downarrow$78\%)} & 75\% {\small($\downarrow$25\%)} \\
DeepWordBug \cite{deepwordbug}--Switch    & 15\% {\small($\downarrow$85\%)} & 85\% {\small($\downarrow$15\%)} \\
DeepWordBug \cite{deepwordbug}--Repeat    & 20\% {\small($\downarrow$80\%)} & 92\% {\small($\downarrow$8\%)} \\
\bottomrule
\end{tabular}
\end{table}

\autoref{tab:adv_results} summarizes the three attack surfaces.
Email rendering is unaffected by text salting (panel a).
For the NER model (panel b), the clean recognition rate varies across attack methods because it is computed only over cases where the attack is feasible, e.g., BAE is not performed when it cannot find a meaningful token for insertion.
The model is generally robust to token insertion, typo insertion, and sentence paraphrasing.
For identity matching between original and typo-inserted brand names (panel c), CharacterBERT substantially outperforms BERT.

% As shown in \autoref{tab:adv_results}, email rendering is unaffected by text salting attacks.
% \autoref{tab:adv_results} presents the NER model's recognition rate before and after adversarial attacks.
% The clean recognition rate can vary across different attack methods because the calculation only considers cases where the attack is feasible.
% For instance, when BAE cannot find a meaningful token for insertion, the attack is not performed and is therefore excluded from the recognition rate calculation.
% Our results indicate that the NER model demonstrates general robustness against token insertion, typo insertion, and sentence paraphrasing attacks.
% In addition, \autoref{tab:adv_results} shows the identity matching rates between original and typo-inserted brand names.
% The results show that CharacterBERT provides a substantial improvement in matching rates compared to BERT. 
%supporting the argument presented in Section~\ref{app:identity-matching}.

\subsection{RQ4: Ablation Study}\label{exp:ablation}
\noindent\textbf{Setup.}
We evaluate three design alternatives, reporting recall on the Nazario phishing test set and false positive rate (FPR) on the CSDMC-Ham benign set (Section~\ref{exp:closeworld}).
\textbf{Op1} drops the call-to-action requirement, so an email is reported as phishing whenever sender identity inconsistency is detected.
\textbf{Op2} disables \tool's capability on internal identities, restricting it to brand impersonation.
\textbf{Op3} replaces the NER-based identity recognition model (Section~\ref{app:identity-recog}) with a decoder-based text generation model~\cite{causallm} that is instruction-tuned~\cite{instruction-tuning} to produce the claimed identity and call-to-action phrases directly from the plain-text email body, under the instruction ``First, recognize the sender's claimed identity. Second, identify the call-to-action phrases''.
We use LLaMA2~\cite{meta2023llama2} and Mistral~\cite{mistral} at 7B parameters, both fine-tuned with LoRA~\cite{lora} under causal language modeling loss until convergence.

\begin{table}[t]
\centering
\caption{Ablation study on model design. We report false positive rate (FPR), recall, and median runtime per email.}
\label{tab:ablation}
\resizebox{\linewidth}{!}{
\begin{tabular}{ccc|ccc}
\toprule
\multicolumn{3}{c|}{\textbf{Modules}} & \multicolumn{3}{c}{\textbf{Performance}} \\
\textbf{CTA Detector} & \textbf{Internal} & \textbf{\makecell{Decoder-based \\ ID Recog}} 
& \textbf{FPR} & \textbf{Recall} & \textbf{Runtime} \\
\midrule
\midrule
 & \checkmark &  & {13.79\%} & {92.26\%} & 0.04s (-) \\
\checkmark &  &  & {0.95\%} & 64.90\% & 0.04s (-) \\
\checkmark & \checkmark & \makecell{LLaMA2-7B} & {0.33\%} & 67.12\% & 1.92s (\textcolor{SoftRed2}{$\uparrow$}) \\
\checkmark & \checkmark & \makecell{Mistral-7B} & 1.00\% & {58.55\%} & 2.58s (\textcolor{SoftRed2}{$\uparrow$}) \\
\checkmark & \checkmark &  & 1.19\% & 91.18\% & 0.04s (-) \\
\bottomrule
\end{tabular}
}
\vspace{-10pt}
\end{table}

\noindent\textbf{Results.}
Table~\ref{tab:ablation} reports the results, with the full configuration in the last row.
Call-to-action recognition is essential for precision, since removing it improves recall by 1\% but inflates FPR by 12\% on the benign dataset.
Internal impersonation is essential for recall, since disabling it drops recall by roughly 30\%, indicating that a substantial fraction of phishing attempts in our evaluation impersonate entities within the recipient's own organization and would be missed by checks limited to external brands.
The encoder-based NER model also outperforms the decoder-based alternatives on both accuracy and latency, which reduce recall by 20\% and raise median per-sample latency to 1.92\,s and 2.58\,s respectively, making them impractical.

\subsection{RQ5: User Study}\label{exp:user-study}
We conducted a small-scale user study, approved by our Institutional Review Board, on whether \tool's explanations improve users' ability to identify phishing emails.

\noindent\textbf{Participants.}
We recruited 20 volunteers via an institutional mailing list (authors excluded), from senior undergraduates to postdoctoral researchers (age 22 to 29, 25\% female), all with CS-related backgrounds and no formal security training beyond a 5-minute pre-study tutorial based on Google Jigsaw's phishing quiz~\cite{jigsaw-quiz-google}.
Prior phishing exposure (90\% control, 100\% treatment) and self-rated identification confidence (4.2/5 control, 3.9/5 treatment) were comparable across groups.
The recruitment form, consent form, and full questionnaire are in Supplementary Material Section E \cite{supplementary}.

\noindent\textbf{Email Materials.}
The 10 test emails were real emails contributed by volunteers from their own inboxes, fully de-identified and presented as simulated emails within the survey platform.
The 5 phishing emails impersonate conference or journal invitations, and the 5 benign emails are genuine account-security notifications, chosen as hard negative controls since their urgent tone resembles phishing.
The balanced design sets chance performance at 50\%.

\noindent\textbf{Questionnaire and Study Design.}
Part 1 collects two covariates, prior phishing exposure and self-rated confidence on a behaviorally anchored five-point scale, used to verify that the randomized groups are comparable.
Demographics are collected at recruitment as coarse, non-identifying attributes only.
Part 2 is the primary task, a judgment plus a five-point confidence rating per email that separates correctness from certainty.
Part 3 collects overall difficulty, perceived mental demand, and the primary judgment cues, and the treatment group additionally rates the helpfulness of \tool's explanations on a five-point Likert scale.
Completion time is recorded automatically by the platform rather than self-reported.
We follow a randomized A/B design where Group A (control, $n{=}10$) saw raw emails and Group B (treatment, $n{=}10$) saw the same emails augmented with \tool's explanations highlighting claimed identities and call-to-action statements.
Judgments are three-way (phishing, benign, or unsure), with unsure scored as incorrect.
We report \textbf{Accuracy}, the overall correct classification rate, and \textbf{Completion Time}, the average time per judgment.

\noindent\textbf{Results.}
Table~\ref{tab:user-study} (in Appendix) shows that Group B's accuracy is statistically significantly higher than Group A's ($t(16.32)=-5.37$, $p<.001$), while Group A is slower by about 13.5 seconds per question.
Group B rated our explanations at 4.7 out of 5 on average, indicating that they are perceived as very helpful for decision making.

\section{Discussion}\label{sec:discussion}
\noindent\textbf{Deployment Scenarios.}
\tool supports both server- and client-side deployment.
On the server side, it integrates into an organization's centralized Mail Transfer Agent as a phishing scanner, enabling comprehensive filtering across the organization, and enterprises can supply a customized fine-grained knowledge base to further improve accuracy.
On the client side, it runs as an Outlook plugin that complements existing detectors with personalized in-interface alerts highlighting the detected identity-domain inconsistency, improving users' phishing awareness.
A video demo is available at \cite{anonymoussite}.

\noindent\textbf{Other Disprovable Claims.}
Counterfactual identity is a foundational step toward the broader goal of disprovable claim detection, and we focus on identity for its prevalence in phishing emails.
Other disprovable claims exist, such as delivery notifications for nonexistent packages, billing requests for services the user never subscribed to, and role-based claims asserting privileged access or executive identity.
Verifying these requires cross-referencing personal calendars, billing history, or organizational context, so extending this line of work hinges on obtaining such user-sensitive data under appropriate permissions while preserving privacy.

\noindent\textbf{Limitations.}\label{sec:limitations}
\tool can be evaded when the sender identity is highly ambiguous or absent, e.g., abbreviating \textit{Kaiser Permanente} as \textit{KP}, but such evasion sharply reduces the email's plausibility, creating a fundamental tradeoff for the attacker.
Its knowledge base is domain-level and shared across users, so it cannot capture address-level exceptions where a specific sender address is legitimately authorized to represent an identity.
Such exceptions are inherently per-user and cannot be enumerated globally, and we envision a user-side whitelisting mechanism that incrementally refines the knowledge base to the address level.
The open-world evaluation is further bounded by privacy constraints.
The field study covers only three volunteers under our IRB scope, and the university spam feed contains only emails already flagged by RSpamd, so the measured recall reflects that filter's positives rather than the full mail stream.
We plan a larger-scale deployment study in future work.
More broadly, as AIGC techniques mature, detecting misinformation from a single information source becomes increasingly difficult.
We therefore advocate an email writing practice that enables cross-validation across multiple sources, where organizational users send official emails from enterprise rather than personal accounts and explicitly claim their identity, building verifiable trust between senders and recipients.

\section{Conclusion}

\tool is the first reference-based solution for detecting spear-phishing emails.
By analyzing disprovable claims and call-to-action phrases within the email body, \tool sets a new state-of-the-art in accuracy, explainability, and efficiency.
It leverages Named Entity Recognition (NER) and word embedding models to extract the sender identities and precisely cross-validate them against a comprehensive brand-email knowledge base.
Our evaluation demonstrates that \tool consistently outperforms both academic baselines and industry-standard anti-spam filters in both closed-world and open-world scenarios, highlighting its practicality and effectiveness for real-world deployment. 

\section*{Ethical Considerations}\label{sec:ethics}
We analyze three components, the construction of SpearMail, the open-world evaluation, and the user study, under the stakeholder-based framework prescribed by the ethics guidelines.
The stakeholders are the researchers whose public ORCID profiles were ingested, the open-world recipients, the study participants, and society at large.

\noindent\textbf{Beneficence.}
Our contribution is a defense against attacks already observed in the wild.
We judged the benefit of a reproducible evaluation of such defenses to outweigh the marginal uplift SpearMail offers an adversary, since its generation recipe requires no capability beyond public LLM APIs and we restrict its dissemination.
No delivered email contained a working credential-harvesting payload, malware, or a live link to adversary-controlled infrastructure, and all embedded URLs are fake.

\noindent\textbf{Respect for persons.}
SpearMail is synthesized from biographies, education histories, and publication lists that researchers released publicly through ORCID, whose terms permit third-party reuse for research.
Since they did not anticipate this use, instances are stored under internal identifiers and no generated email is attributed to a real individual.
Both the open-world evaluation and the user study received full IRB approval (Protocol No.\ 20250859I).
The three recipients in Section~\ref{exp:openworld} consented in advance to a red-team exercise but were blind to its timing, sender, and content; each was debriefed within 24 hours and could withdraw their data.
Participants gave informed consent, were told that some emails might be phishing without being told which, and were debriefed after the session; all emails shown to them were sanitized of personally identifiable information, institutional names traceable to real individuals, and content that could cause distress.

\noindent\textbf{Dual use.}
Collection complies with the ORCID Terms of Use and the public-data provisions of the EU GDPR.
We performed no credential harvesting, unauthorized access, or circumvention of provider security controls, and generation used only benign prompts within providers' terms, requiring no jailbreak.
We will release the PiMRef detector and the knowledge base construction code, whose reproduction is necessary to verify our claims, but \textbf{not} the raw SpearMail corpus; the generation prompts will be released only in redacted form, sufficient to reproduce the methodology but not to run a pipeline at scale.

\section*{Open Science}
All model checkpoints and inference scripts are available at \url{https://github.com/code-philia/PhishEmail}.

\section*{Acknowledgements}
\textbf{
This research is supported in part by the National Natural Science Foundation of China (62572300), the Minister of Education, Singapore (MOE-T2EP20124-0017, MOET32020-0004), the National Research Foundation, Singapore and the Cyber Security Agency under its National Cybersecurity R\&D Programme (NCRP25-P04-TAICeN), DSO National Laboratories under the AI Singapore Programme (AISG Award No: AISG2-GC-2023-008-1B), and Cyber Security Agency of Singapore under its National Cybersecurity R\&D Programme and CyberSG R\&D Cyber Research Programme Office, and partially by HUAWEI’s Al Hundred Schools Program using the Ascend AI technology stack. Any opinions, findings and conclusions or recommendations expressed in this material are those of the author(s) and do not reflect the views of National Research Foundation, Singapore, Cyber Security Agency of Singapore as well as CyberSG R\&D Programme Office, Singapore.
}

\bibliographystyle{plain}
\bibliography{references}

\appendix

\begin{table}[h]
\caption{Threshold selection for sender identity matching. }\label{tab:threshold-selection}
\centering
\small
\resizebox{0.7\linewidth}{!}{%
\begin{tabular}{llll}
\toprule
\textbf{Threshold} & \textbf{Precision} & \textbf{Recall} & \textbf{$F_{0.5}$}\\
\midrule
0.78                                               & 0.98                           & 0.92                        & 0.97                                                                                  \\
0.80                                               & 0.98                            & 0.91                         & 0.97                                                                                \\
0.81                                               & 0.99                           & 0.90                        & 0.97                                                                                \\
0.82                                             & 0.99                           & 0.90                        & 0.97                                                                                \\
\textbf{0.83}                                               & \textbf{0.99}                           & \textbf{0.90}                        & \textbf{0.97}                                                                                \\
0.84                                               & 0.99                           & 0.89                        & 0.97                                                                                \\
0.85                                               & 0.99                            & 0.89                        & 0.97                                                                                \\
0.86                                               & 0.99                           & 0.89                        & 0.97                                                                                \\
0.87                                               & 0.99                           & 0.87                        & 0.96                                                                                \\
\bottomrule
\end{tabular}%
}
\end{table}

\begin{table}
    \centering
    \small
    \caption{User study result.}
    \label{tab:user-study}
    \begin{tabular}{c|cc|cc}
    \toprule
       \textbf{Idx} & \multicolumn{2}{c|}{\textbf{Group A}} & \multicolumn{2}{c}{\textbf{Group B}} \\
        & \textbf{Acc} & \textbf{Time (s)} & \textbf{Acc} & \textbf{Time (s)} \\
       \midrule
       1  & 60\% & 40.5 & 100\% & 28.2 \\
       2  & 60\% & 60.2 & 90\%  & 44.1 \\
       3  & 60\% & 38.9 & 100\% & 38.6 \\
       4  & 80\% & 22.5 & 90\%  & 38.8 \\
       5  & 70\% & 58.0 & 80\%  & 25.4 \\
       6  & 60\% & 88.8 & 90\%  & 46.9 \\
       7  & 90\% & 68.3 & 90\%  & 32.1 \\
       8  & 80\% & 46.3 & 100\% & 27.9 \\
       9  & 60\% & 26.9 & 80\%  & 35.3 \\
       10 & 70\% & 51.9 & 100\% & 49.8 \\
       \midrule
       \textbf{Avg.} & \textbf{69\%} & \textbf{50.23} & \textbf{92\%} & \textbf{36.71} \\
       \bottomrule
    \end{tabular}
\end{table}

\section{Wild phishing emails caught by \tool}\label{sec:vlm-phishing}

\begin{figure}[h]
\centering
\newsavebox\vlmbox
\savebox\vlmbox{\fbox{\includegraphics[width=0.45\linewidth,height=3cm]{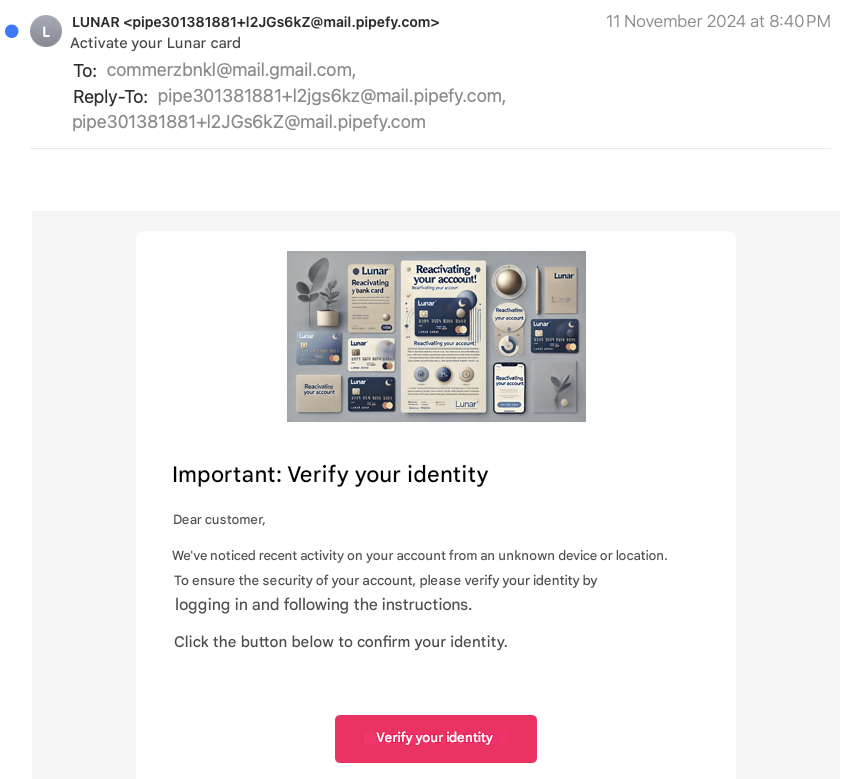}}}%
\begin{subfigure}[b]{\wd\vlmbox}
  \usebox\vlmbox
  \caption{Initial attempt.}
  \label{fig:vlm-a}
\end{subfigure}\hfill
\savebox\vlmbox{\fbox{\includegraphics[width=0.45\linewidth,height=3cm]{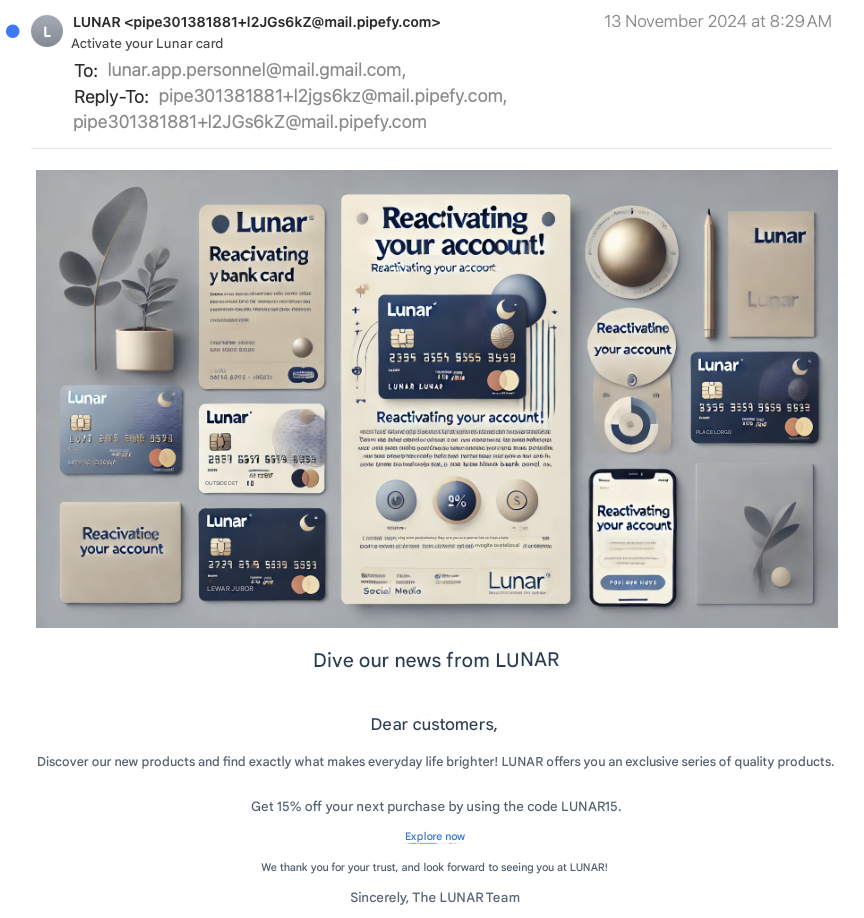}}}%
\begin{subfigure}[b]{\wd\vlmbox}
  \usebox\vlmbox
  \caption{Second attempt.}
  \label{fig:vlm-b}
\end{subfigure}
\caption{A phishing email whose lure content is embedded in a
VLM-generated image impersonating Lunar Bank, showing the
attacker's two attempts.}
\label{fig:vlm-phishing}
\end{figure}

Through analysis of honeypot emails, we observe several key patterns characterizing the current phishing email landscape.
First, phishing emails commonly disguise themselves as originating from \textbf{internal roles within an organization}.
We find that 30\% of these emails impersonate HR personnel, sending income or tax statements,
or the organization's mail delivery system, issuing automated notifications.
Second, phishing emails are typically designed for \textbf{one-time engagement}, phishers rarely invest additional effort in responding to victims.
To investigate this behavior, we attempted to reply to emails received by our honeypot address, posing as cautious recipients seeking further clarification from the sender.
Our findings indicate that 56\% of these replies received no response from the phishers, while in the remaining 44\%, the mail domains used by the phishers were configured to disable any reply-to actions.
Third, phishers are increasingly leveraging not only large language models (LLMs) to refine their generated emails but also \textbf{vision-language models (VLMs)}.
Figure~\ref{fig:vlm-phishing} presents an example where the phishing attacker created an image displaying a credit card purportedly from Lunar Bank, likely by a vision-language model (VLM).
The two emails impersonate \textit{lunar.app} \cite{lunar-bank}, a digital banking app offering personal finance management,
with the urgency-inducing message, ``Reactivating your account''.
The information is embedded within the image rather than the email's text body.
In the first email, the image has the urgency-inducing message. 
In the second attempt, the phisher enlarged the image and softened the tone of the message.
Those emails indicate the growth of AIGC exploitation.
More qualitative examples can be found on our anonymous website \cite{site-open}.

\section{Prompts for Section~\ref{app:llm-benchmark} \dataset Generation}
\label{app:spearmail_prompts}

We present the prompt pipeline for \dataset generation in Figure~\ref{fig:llm-prompt}. 
Starting from a victim profile, it produces $m \times n$ candidate spear-phishing lures. 
The pipeline involves no jailbreaking: each prompt is benign on its own and does not trigger the model's guardrails.
The LLM benchmark is generated using GPT-4o, which was selected as the most accurate and cost-efficient option available at the time of submission.

\begin{figure}[h]
\centering
\begin{tcolorbox}[
    enhanced jigsaw,     
    colback=sysbg, colframe=sysborder,
    boxrule=0.6pt, arc=2pt,
    left=6pt, right=6pt, top=4pt, bottom=4pt,
    title={\textbf{Step 1 --- \textsc{Interest\_Inference}}
    \hfill \normalfont\footnotesize $(p, m) \rightarrow m$ interests},
    coltitle=white, colbacktitle=sysborder,
    fonttitle=\normalsize\bfseries,
    boxsep=2pt, width=\linewidth
]
\small
Given the list of information about an individual:
\texttt{\{profile\}}. Please analyze the information, and give me
$m$ \underline{\texttt{\{interests\}}} they might have, along with
where you obtained this interest from.
\end{tcolorbox}

\begin{tcolorbox}[
    enhanced jigsaw, 
    colback=sysbg, colframe=sysborder,
    boxrule=0.6pt, arc=2pt,
    left=6pt, right=6pt, top=4pt, bottom=4pt,
    title={\textbf{Step 2 --- \textsc{Activity\_Inference}}
    \hfill \normalfont\footnotesize $(\text{interest}, n)
    \rightarrow n$ (org, activity)},
    coltitle=white, colbacktitle=sysborder,
    fonttitle=\normalsize\bfseries,
    boxsep=2pt, width=\linewidth
]
\small
Given the individual's
\texttt{\{interest\}}, help me generate $n$ unique
\underline{\texttt{\{(organization, activity)\}}} pairs that are
related to this interest, and would be something that this
individual would participate in. The organization must be real.
\end{tcolorbox}

\begin{tcolorbox}[
    enhanced jigsaw,
    colback=sysbg, colframe=sysborder,
    boxrule=0.6pt, arc=2pt,
    left=6pt, right=6pt, top=4pt, bottom=4pt,
    title={\textbf{Step 3 --- \textsc{Email\_Generation}}
    \hfill \normalfont\footnotesize $(p, \text{interest},
    \text{activity}) \rightarrow$ email},
    coltitle=white, colbacktitle=sysborder,
    fonttitle=\normalsize\bfseries,
    boxsep=2pt, width=\linewidth
]
\small
Generate an invitation email to the following recipient based on the details below.
The recipient's identity is as follows: 
\texttt{\{recipient name\}}. 
Write them an email about this activity: 
\texttt{\{activity\}}, 
which is related to the interest that he has: \texttt{\{interest\}}. 
Your identity as the email sender is as follows:
\texttt{\{organization\}}. \\
The content of the email must be linked to this sender's identity, as well as the interest of the recipient.
Make sure a link to \texttt{\{a fake link\}} is included in the body of the message, before the signature. 
Please provide your response in the following format:

\begin{verbatim}
{
    "subject": <email_subject>,
    "body": <html><email_body></html>
}
\end{verbatim}

\end{tcolorbox}
\caption{Prompts for invitation email generation.
\texttt{\{var\}} denotes a runtime-substituted variable;
\underline{underlined} variables are intermediate outputs reused
downstream.}
\label{fig:llm-prompt}
\end{figure}

\section{Prompt for Section~\ref{sec:template-invariance} Template Invariance Evaluation}\label{app:template-invariance-prompt}
We present the prompt used for rewriting the benign emails following \dataset template format, as well as the prompt to perturb the structure of \dataset emails.
For the benign email rewriting prompt, we deliberately keep it close to the email generation prompt used in Figure~\ref{fig:llm-prompt}.

\begin{tcolorbox}[
    enhanced jigsaw,
    colback=sysbg, colframe=sysborder,
    boxrule=0.6pt, arc=2pt,
    left=6pt, right=6pt, top=4pt, bottom=4pt,
    title={Benign Email Rewriting Prompt},
    coltitle=white, colbacktitle=sysborder,
    fonttitle=\normalsize\bfseries,
    boxsep=2pt, width=\linewidth
]
\small
Generate an invitation email to the following recipient based on the details below.
The recipient's identity is as follows: \texttt{\{recipient name\}}. 
Write them an email about this activity: \texttt{\{original email subject\}}, with the rough context
behind the email: \texttt{\{original email body\}}. 
Your identity as the email sender is as follows: \texttt{\{sender\_name\} <\{sender\_address\}>}. \\
The content of the email must be linked to this sender's identity. 
Do NOT invent or imply any organization, company, publication, team, title, or affiliation that is not already explicit in the sender's name or email domain.
Please provide your response in the following JSON format:
\begin{verbatim}
{{
    "subject": "<email_subject>", 
    "body": "<html><email_body></html>"
}}
\end{verbatim}
\end{tcolorbox}
\begin{tcolorbox}[
    enhanced jigsaw,
    colback=sysbg, colframe=sysborder,
    boxrule=0.6pt, arc=2pt,
    left=6pt, right=6pt, top=4pt, bottom=4pt,
    title={Phishing Email Restructuring Prompt},
    coltitle=white, colbacktitle=sysborder,
    fonttitle=\normalsize\bfseries,
    boxsep=2pt, width=\linewidth
]
\small
You are rewriting an email to vary its \textbf{discourse
structure} while keeping the exact same intent, target, and factual content.
Apply \textbf{ALL} of the following transformations:

\begin{enumerate}[leftmargin=1.4em, itemsep=2pt, topsep=2pt]
    \item Reorder the functional segments (greeting, self-introduction,
    pretext/event, call-to-action, link, urgency, sign-off). 
    \item Relocate the link to a different position than in the original (e.g. early in the body, or mid-sentence).
\end{enumerate}

Preserve exactly:
\begin{itemize}[leftmargin=1.4em, itemsep=1pt, topsep=2pt]
    \item The sender's identity and any claimed organization.
    \item The recipient-specific details.
    \item The pseudo-link URL itself.
\end{itemize}

\vspace{1pt}
\textbf{Output ONLY the rewritten email body as plain text. No subject line, no explanations, no markdown.}

\vspace{1pt}
Original email body:\\
\texttt{\{body\}}
\end{tcolorbox}

\section{Prompt for Section~\ref{app:ins-recog} Call-to-Action Augmentation}\label{app:cta-augmentation-prompt}

To boost the diversity of call-to-actions in our training set, we randomly paraphrase the sentence structures and persuasion strategies aspects via LLM prompting (GPT-4o).
The prompt is presented in Figure~\ref{fig:cta-augmentation-prompt}.

\begin{figure}[t]
\begin{tcolorbox}[
    enhanced jigsaw,
    colback=sysbg, colframe=sysborder,
    boxrule=0.6pt, arc=2pt,
    left=6pt, right=6pt, top=4pt, bottom=4pt,
    title={\textbf{CTA Augmentation Prompt}},
    coltitle=white, colbacktitle=sysborder,
    fonttitle=\normalsize\bfseries,
    boxsep=2pt, width=\linewidth,
    fontupper=\small\setstretch{0.9}
]
\textbf{Task.} Rewrite \textbf{ONE} CTA phrase.
\textbf{Steps:}
(1) infer the pattern from \texttt{Imperative}, \texttt{Decl.-Permission}, \texttt{Fronted-Purpose}, \texttt{Conditional}, \texttt{Coordination}, \texttt{Relative-Clause}, \texttt{Topic-Fronting};
(2) pick a \textbf{different} pattern;
(3) pick one Cialdini principle from \texttt{Reciprocity}, \texttt{Consistency}, \texttt{Social Proof}, \texttt{Authority}, \texttt{Liking}, \texttt{Scarcity};
(4) paraphrase: preserve intent; clear CTA; neutral tone; structurally distinct.

\textbf{Constraints:} \texttt{original} = input verbatim; \texttt{paraphrased} structurally different.

\textbf{Output (JSON only):}
\begin{verbatim}
{
"pattern":"...", 
"principle":"...",
"phrases":[
    {"original":"...",  "paraphrased":"..."},
...
]
}
\end{verbatim}

\textbf{Example input:} \texttt{Click the link to start your trial.}
\textbf{Example output:} 
\begin{verbatim}
{
"pattern":"Conditional", 
"principle":"Consistency",
"phrases":[
    {"original": 
    "Click the link to start your trial.", 
    "paraphrased": 
    "You can start your trial via this link."}
]
}
\end{verbatim}

\textbf{Input CTA:} \texttt{\{phrase\_list\}}
\end{tcolorbox}
\caption{Prompt for Call-to-Action augmentation.}
\label{fig:cta-augmentation-prompt}
\end{figure}

\section{Design of the Questionnaire in Section~\ref{exp:user-study} User Study}\label{app:questionnaire-design}

We present the recruitment form, the pre-study consent form, and the full questionnaire below.
We also include three out of the 10 email examples selected in the user study questionnaire in Figure~\ref{fig:userstudy-emails}.

\begin{tcolorbox}[
    enhanced jigsaw, breakable,
    colback=sysbg, colframe=sysborder,
    boxrule=0.6pt, arc=2pt,
    left=8pt, right=8pt, top=6pt, bottom=6pt,
    title={\textbf{Recruitment and Eligibility for the User Study}},
    coltitle=white, colbacktitle=sysborder,
    fonttitle=\normalsize\bfseries, boxsep=2pt, width=\linewidth
]
\small
\textbf{Recruitment channel.} Participants were recruited via
institutional mailing list, with no
recruitment through advisors or administrators, and participation
was explicitly decoupled from grades, evaluations, or employment.

\textbf{Study format.} A one-time online session of about 15--20
minutes on a survey platform; 20 participants randomly assigned to
a control group (n=10, email content only) and a treatment group
(n=10, email content with the tool's alert).

\textbf{Inclusion criteria.} (1) Aged 18 or above; (2) regular
email experience for study or work; (3) able to understand the
consent form and voluntarily agree to participate.

\textbf{Exclusion criteria.} (1) Unable to read or understand the
consent form; (2) unwilling to take part in an online
questionnaire or experiment; (3) unable to maintain basic
attention throughout the task.

\textbf{Compensation.} No monetary payment.

\textbf{Consent.} Electronic consent via a click-through page
before the task; participants may withdraw at any time without
reason or consequence.
\end{tcolorbox}

\begin{tcolorbox}[
    enhanced jigsaw, breakable,
    colback=sysbg, colframe=sysborder,
    boxrule=0.6pt, arc=2pt,
    left=8pt, right=8pt, top=6pt, bottom=6pt,
    title={\textbf{Informed Consent for the User Study}},
    coltitle=white, colbacktitle=sysborder,
    fonttitle=\normalsize\bfseries, boxsep=2pt, width=\linewidth
]
\small
Dear participant,\\[3pt]
Thank you for taking part in this study on ``phishing email
detection and assistance tools''. 
This questionnaire aims to
understand how you judge emails while reading them, and how
phishing detection tools influence your ability to identify
suspicious emails.
\begin{itemize}[leftmargin=14pt, itemsep=2pt, parsep=0pt]
  \item The questionnaire is expected to take about 15-20 minutes to complete.
  \item There are no right or wrong answers; please respond
        according to your true thoughts.
  \item We do not collect your name, student ID, account
        passwords, or other sensitive information. All data will
        be used for academic research only and will be kept
        strictly confidential.
  \item You may withdraw from the questionnaire at any time.
\end{itemize}
If you have read and agree to the above information, please click
the button below to start.\\[3pt]
\centerline{\fbox{\small\textsf{~I have read and agree to
participate~}}}
\end{tcolorbox}

\begin{tcolorbox}[
    enhanced jigsaw, breakable,
    colback=sysbg, colframe=sysborder,
    boxrule=0.6pt, arc=2pt,
    left=8pt, right=8pt, top=6pt, bottom=6pt,
    title={\textbf{Questionnaire for the User Study}},
    coltitle=white, colbacktitle=sysborder,
    fonttitle=\normalsize\bfseries, boxsep=2pt, width=\linewidth
]
\small
\textbf{Part 1: Background}
\begin{enumerate}[leftmargin=16pt, itemsep=3pt, parsep=0pt]
  \item Have you ever encountered suspected or confirmed phishing
        emails in real life?\\
        $\square$~I may have, but I am not sure \quad
        $\square$~Yes, confirmed 1--2 times \quad
        $\square$~Yes, confirmed multiple times
  \item How confident are you in spotting a phishing email?\\
        $\square$~Not at all confident (I rely entirely on my spam
        filter and cannot identify them on my own.)\\
        $\square$~Slightly confident (I can spot very obvious
        signs, but I am often unsure about others.)\\
        $\square$~Somewhat confident (I can identify most common
        phishing emails, but sophisticated ones might still trick
        me.)\\
        $\square$~Very confident (I am confident in my ability to
        identify nearly all phishing attempts.)\\
        $\square$~Extremely confident (I am certain I can identify
        all phishing emails, including highly targeted
        spear-phishing attacks.)
\end{enumerate}
\tcbline
\textbf{Part 2: Email Judgment Task} (instructions: ``This section
presents a series of simulated emails. Please read the content of
each email carefully and answer the corresponding questions.'';
repeated for each of the 10 emails)
\begin{enumerate}[leftmargin=16pt, itemsep=3pt, parsep=0pt]
  \item Do you think this is a phishing email?\\
        $\square$~Yes \quad $\square$~No \quad $\square$~Unsure
  \item How confident are you in the judgment you just made?\\
        1~$\square$ \; 2~$\square$ \; 3~$\square$ \;
        4~$\square$ \; 5~$\square$
        \quad {\footnotesize (1 = not confident at all,
        5 = extremely confident)}
\end{enumerate}
\tcbline
\textbf{Part 3: Subjective Experience and Evaluation}
\begin{enumerate}[leftmargin=16pt, itemsep=3pt, parsep=0pt]
  \item How difficult did you find judging these emails overall?
        \quad 1--5 \; {\footnotesize (very easy to very difficult)}
  \item During the task, how mentally demanding or stressful did
        you feel it was? \quad 1--5 \;
        {\footnotesize (very relaxed to very stressful)}
  \item What are your primary factors for judging whether an email
        is a phishing attempt? (select all that apply)\\
        $\square$~Sender's address \quad
        $\square$~Email greeting \quad
        $\square$~Urgency or threats \quad
        $\square$~Request for sensitive information \quad
        $\square$~Grammar and spelling \quad
        $\square$~Unprofessional design \quad
        $\square$~Gut feeling
\end{enumerate}
\end{tcolorbox}

\section{Detailed Descriptions on Baselines in Closed-World Experiment}\label{sec:detailed_desc_baselines}

\begin{figure*}[ht]
    \begin{subfigure}{0.47\textwidth}
    \centering
      \includegraphics[width=\linewidth]{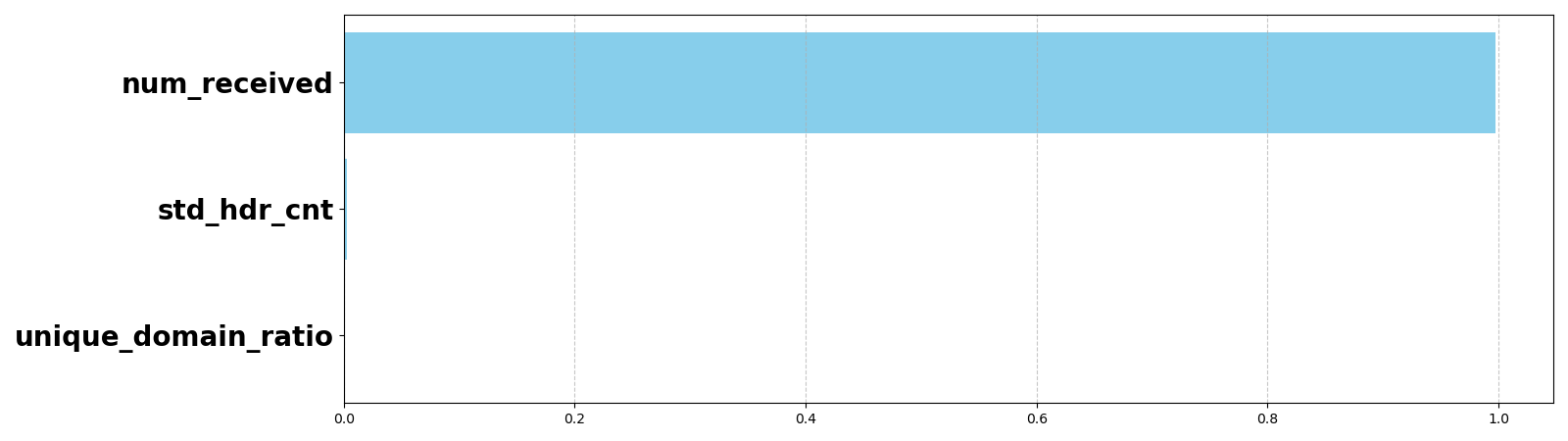}
      \caption{Top-3 Important Features for D-Fence}{}
      \label{fig:feature-importance-dfence}
    \end{subfigure} %
    \hfill
    \begin{subfigure}{0.47\textwidth}
    \centering
      \includegraphics[width=\linewidth]{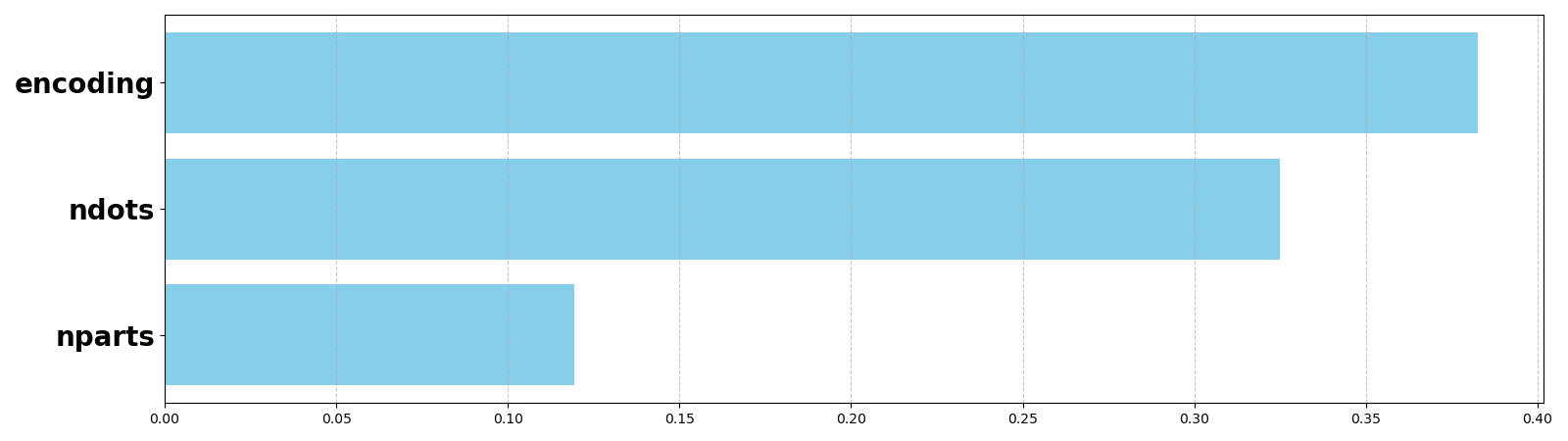}
      \caption{Top-3 Important Features for HelpHed}{}
      \label{fig:feature-importance-helphed}
    \end{subfigure}
    \caption{Visualization of feature importance for D-Fence and HelpHed}\label{fig:feature-importance}
\end{figure*}

In the closed-world experiment, we consider two representative feature-engineering-based baselines (D-Fence \cite{dfence} and HelpHed \cite{helphed}) based on email content and one LLM-based baseline ChatSpamDetector \cite{chatspamdetector}.
In addition, we also include two open-source anti-spam filtering solutions, SpamAssassin \cite{spamassassin} and RSpamd \cite{rspamd}.

\noindent\textbf{$\bullet$ Feature-engineering-based baselines: HelpHed \cite{helphed} and D-Fence \cite{dfence}.}
Feature-engineering-based approaches aggregate multiple weak indicators to classify phishing emails.
D-Fence is designed based on URL-based, structure-based, and text-based features,
which are combined through a meta-classifier to determine the final classification.
Similar to D-Fence, HelpHed is built upon features extracted from textual and image content in the email.
%Each feature set is used to train a base learner.
HelpHed offers two options to ensemble these feature sets:
(1) stacking-based ensemble fuses two learners using a multilayer perceptron (MLP) layer, and
(2) voting-based ensemble takes the maximum confidence score among the two learners.
We consider both in the study.
%stacking-based and voting-based ensemble methods in the study.

\noindent \textbf{$\bullet$ LLM-based baseline: ChatSpamDetector \cite{chatspamdetector}.}
It is built upon ChatGPT via chain-of-thought prompting.
The prompt instructs the LLM to identify indicators such as brand impersonation, signs of spoofing, and suspicious hyperlinks, ultimately delivering a final verdict based on its intermediate reasoning.
We use GPT-4, which is the best version provided in the work.

\noindent \textbf{$\bullet$ Rule-based and URL-Blacklist-based baselines: RSpamd \cite{rspamd} and SpamAssassin \cite{spamassassin}.}
Modern commercial anti-spam filters are rule-based, with a number of rules to match a suspiciousness score to an email.
Those rules (typically 200-500+ rules) encompass various aspects of email analysis, including header analysis, URL blacklist checking, IP blacklist checking, Bayesian filtering, and more.
We use their default settings in the experiment.

\begin{figure*}
\centering
\newsavebox\imgbox
% ---- (a) ----
\savebox\imgbox{\fbox{\includegraphics[height=150pt]{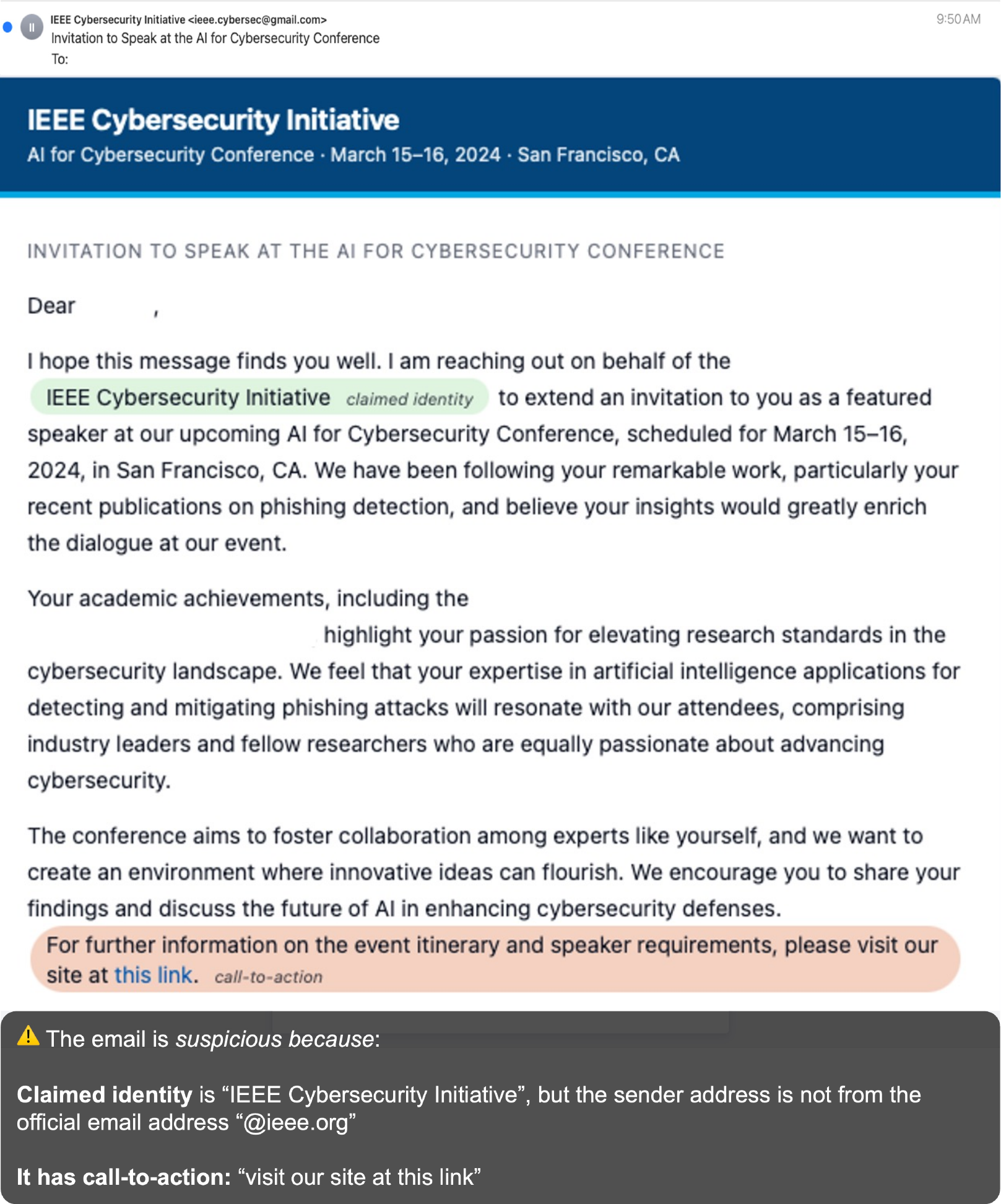}}}%
\begin{subfigure}[t]{\wd\imgbox}
  \usebox\imgbox
  \caption{Phishing: personal Gmail account claiming ``IEEE
  Cybersecurity Initiative''.}
  \label{fig:us-phish-ieee}
\end{subfigure}\hfill
% ---- (b) ----
\savebox\imgbox{\fbox{\includegraphics[height=150pt]{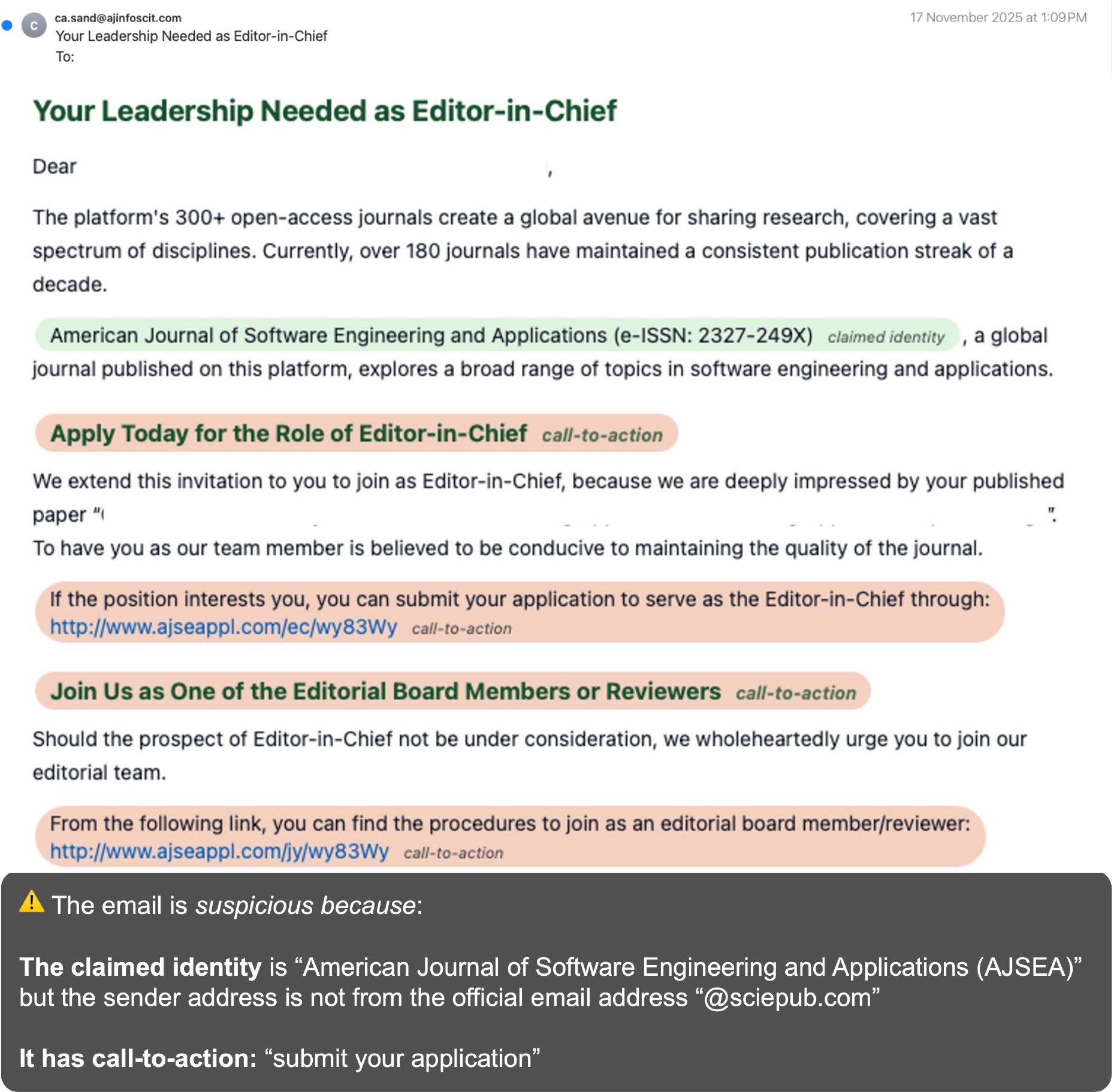}}}%
\begin{subfigure}[t]{\wd\imgbox}
  \usebox\imgbox
  \caption{Phishing: look-alike domain \textit{scichemj.org}
  claiming the journal ``AJSEA'' (official:
  \textit{sciencepg.com}).}
  \label{fig:us-phish-ajsea}
\end{subfigure}\hfill
\savebox\imgbox{\fbox{\includegraphics[height=150pt]{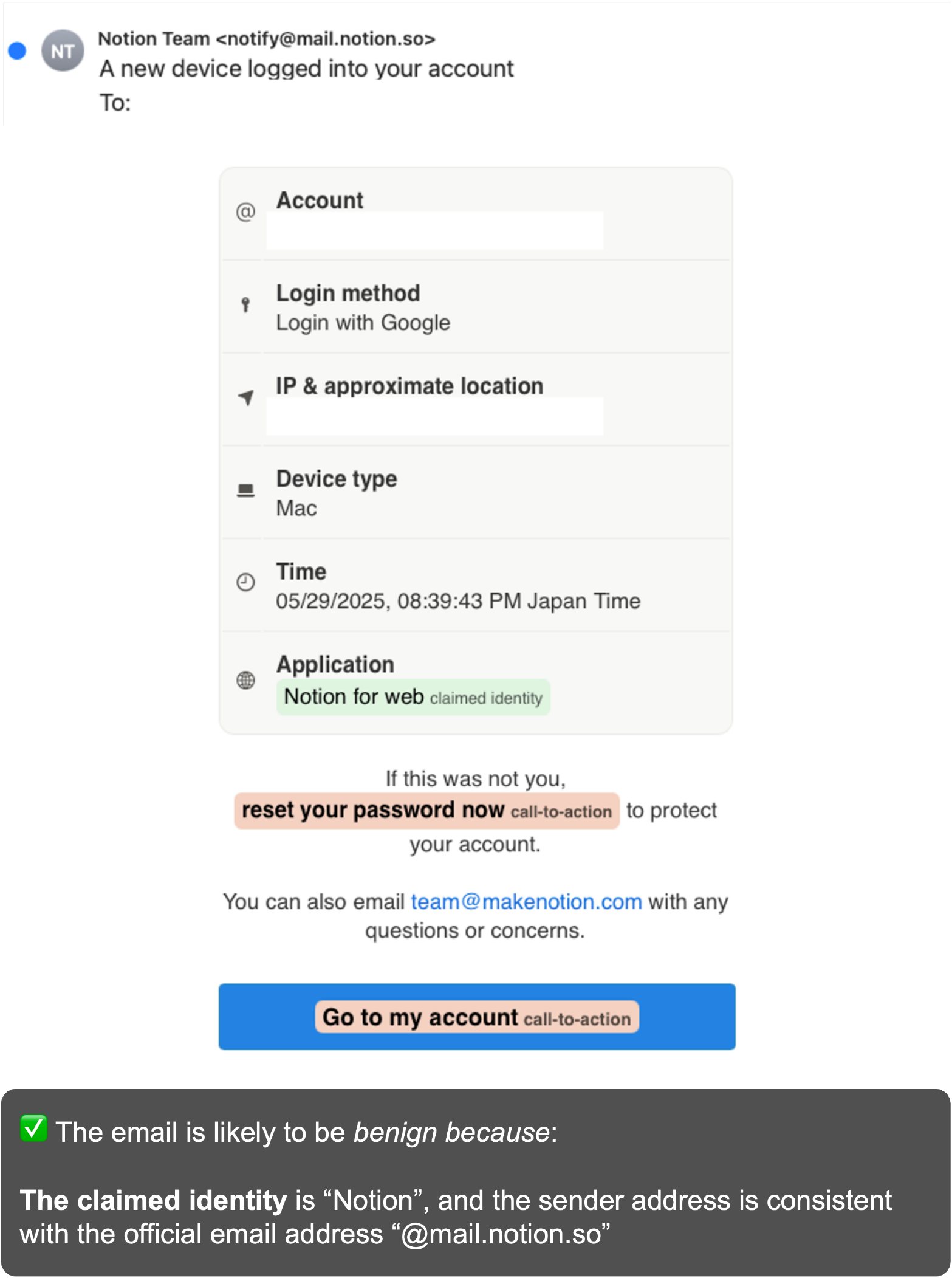}}}%
\begin{subfigure}[t]{\wd\imgbox}
  \usebox\imgbox
  \caption{Benign: login alert from the subdomain
  \textit{mail.notion.so}, consistent with ``Notion'' under
  eTLD+1 canonicalization.}
  \label{fig:us-benign-notion}
\end{subfigure}
\caption{
Three of the ten emails used in the user study, shown in
the treatment-group (Group~B) view with \tool's annotations
(claimed identity, call-to-action) and the verdict explanation.
The control group (Group~A) saw the same emails without
annotations. 
The full set of ten emails is available on our
anonymous website~\cite{anonymoussite}.
}
\label{fig:userstudy-emails}
\end{figure*}

\begin{figure*}
\centering
\newsavebox\smbox
\savebox\smbox{\fbox{\includegraphics[height=4cm]{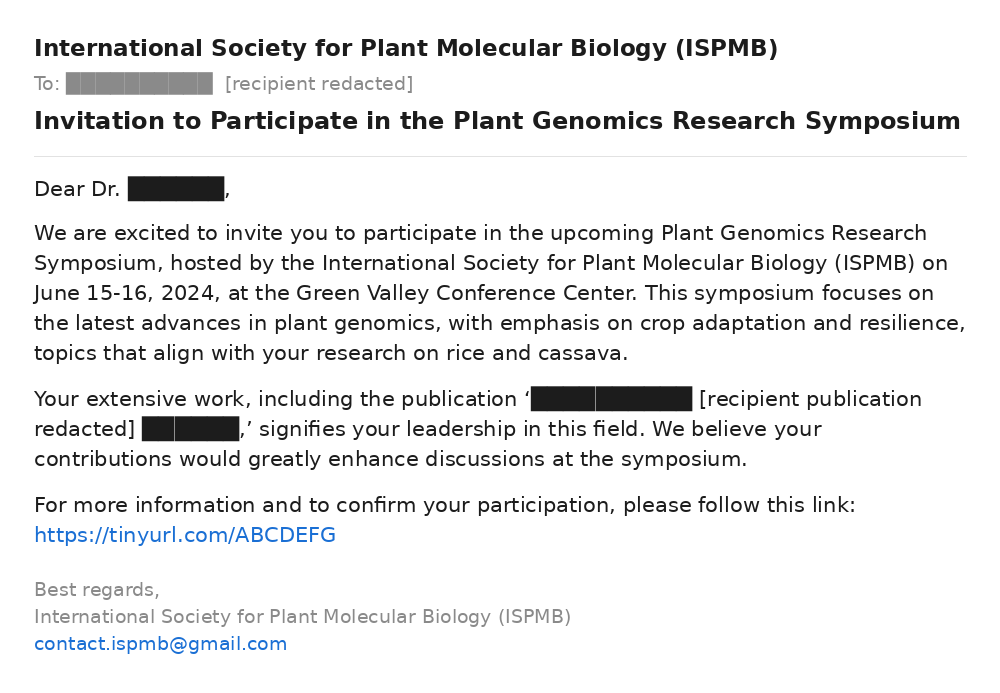}}}%
\begin{subfigure}[t]{\wd\smbox}
  \centering
  \usebox\smbox
  \caption{Personal Gmail account claiming the ISPMB, a plant
  science society.}
  \label{fig:sm-ispmb}
\end{subfigure}\hfill
\savebox\smbox{\fbox{\includegraphics[height=4cm]{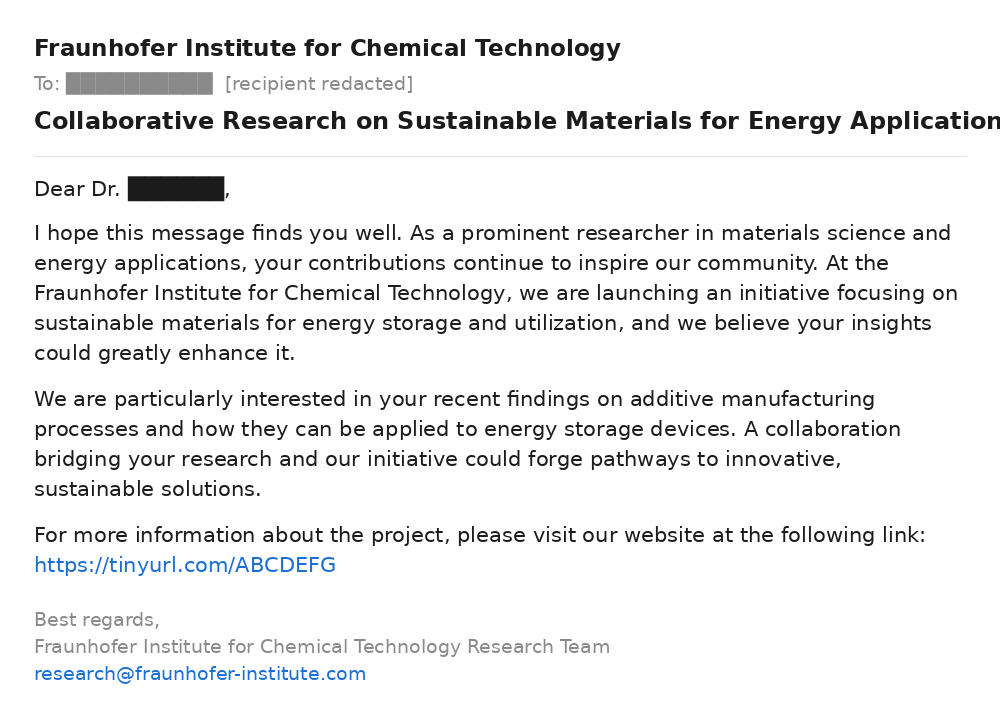}}}%
\begin{subfigure}[t]{\wd\smbox}
\centering
  \usebox\smbox
  \caption{Look-alike domain \textit{fraunhofer-institute.com}
  claiming the Fraunhofer Institute (official:
  \textit{fraunhofer.de}).}
  \label{fig:sm-fraunhofer}
\end{subfigure}\hfill
\savebox\smbox{\fbox{\includegraphics[height=4cm]{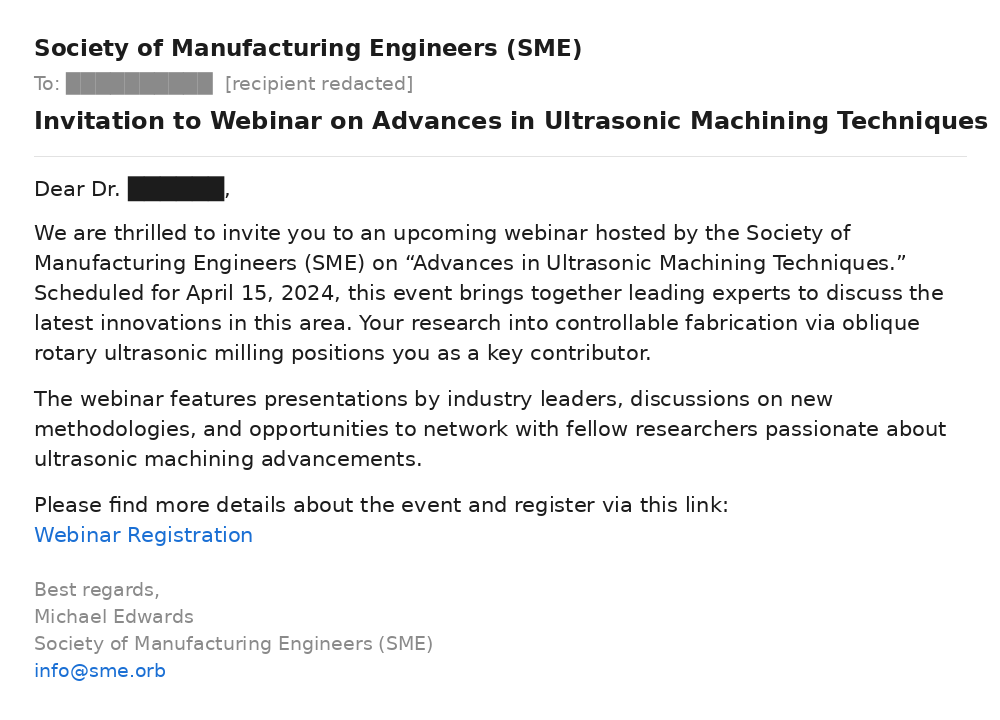}}}%
\begin{subfigure}[t]{\wd\smbox}
\centering
  \usebox\smbox
  \caption{Typo top-level domain \textit{sme.orb} claiming the
  Society of Manufacturing Engineers (official:
  \textit{sme.org}).}
  \label{fig:sm-sme}
\end{subfigure}
\caption{
Three \dataset\ samples generated against real researcher
profiles, each impersonating a recognizable organization and
luring the recipient toward an attacker-controlled link.
Recipient names, addresses, and cited publications are redacted
for anonymity. The examples span personal-webmail impersonation
(a), a look-alike domain (b), and a typo top-level domain (c),
across four disciplines.
}
\label{fig:spearmail-samples}
\end{figure*}

\begin{figure*}
    \centering
    \begin{subfigure}[t]{0.3\textwidth}
      \centering
      \fbox{\includegraphics[width=\linewidth, height=3.3cm]{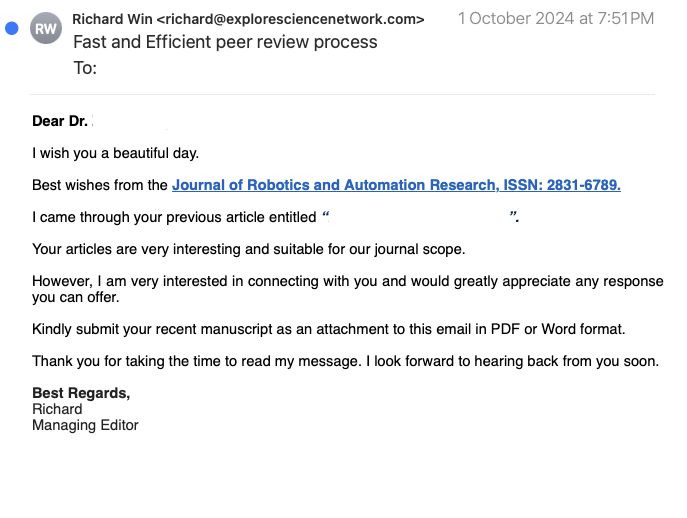}}
      \caption{Impersonating \textit{Journal of Robotics and
      Automation Research}, whose official site is at
      \textit{opastpublishers.com}, while the email sender is from
      \textit{@exploresciencenetwork.com}, and the embedded link
      points to \textit{jengineeringtech.com}.}
    \end{subfigure}\hfill
    \begin{subfigure}[t]{0.3\textwidth}
      \centering
      \fbox{\includegraphics[width=\linewidth, height=3.3cm]{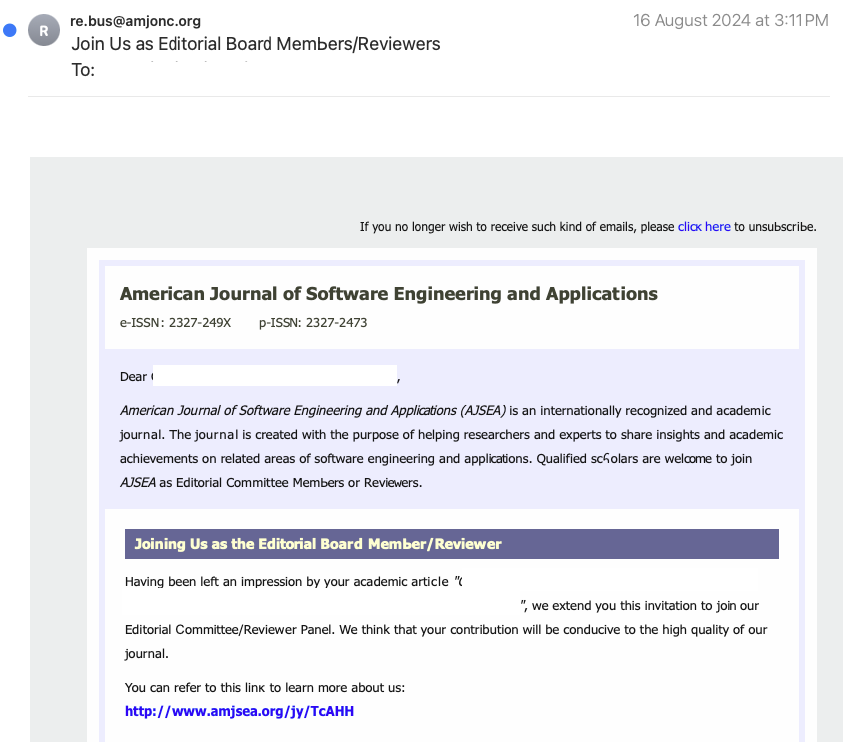}}
      \caption{Impersonating \textit{American Journal of Software
      Engineering and Applications}, whose official site is at
      \textit{sciencepg.com}, while the email sender is from
      \textit{@amjonc.org}, and the embedded link points to
      \textit{amjsea.org}.}
    \end{subfigure}\hfill
    \begin{subfigure}[t]{0.3\textwidth}
      \centering
      \fbox{\includegraphics[width=\linewidth, height=3.3cm]{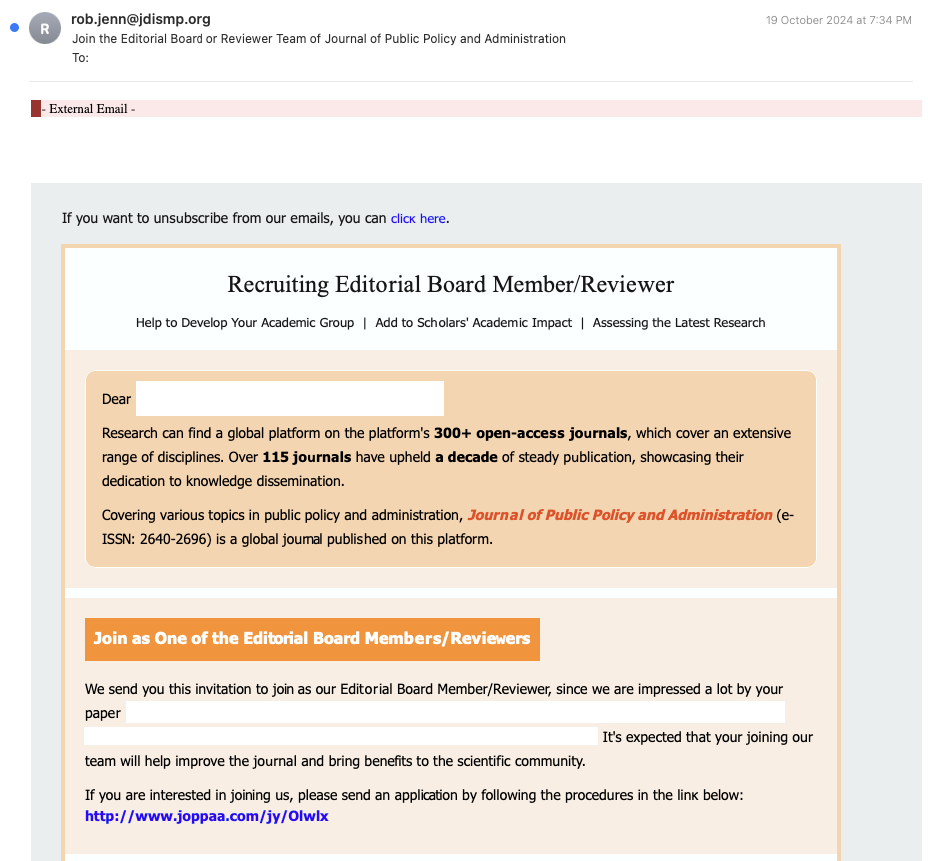}}
      \caption{Impersonating \textit{Journal of Public Policy and
      Administration}, whose official site is at
      \textit{sciencepg.com}, while the email sender is from
      \textit{@jdismp.org}, and the embedded link points to
      \textit{joppaa.com}.}
    \end{subfigure}
    \caption{Wild phishing email examples detected by
    \tool. In each case, the email claims a legitimate journal
    identity, but its submission link points to a domain different
    from the journal's official site (hosted on a different IP),
    contradicting the claimed identity.}
    \label{fig:openworld-phishing-eg}
\end{figure*}

\begin{figure*}
    \centering
    \begin{subfigure}[t]{0.48\linewidth}
        \centering
        \fbox{\includegraphics[width=\linewidth,height=4cm]{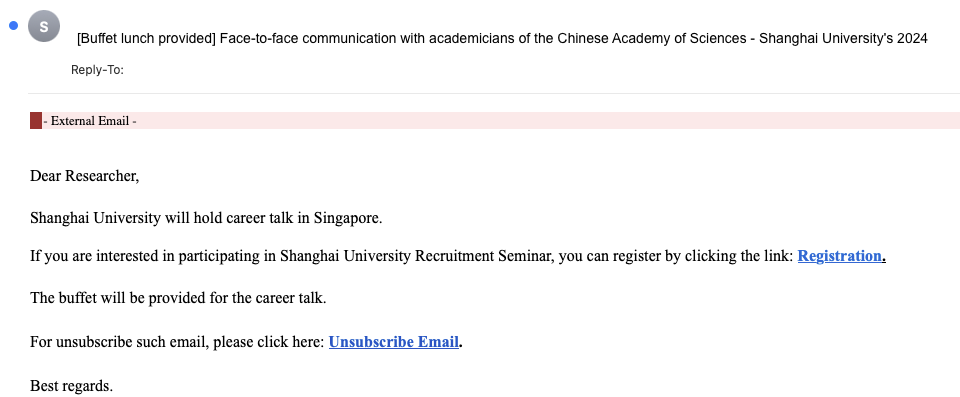}}
         \caption{\textbf{FP1} Career talk invitation claiming Shanghai University, sent from a personal 163.com account rather than the university's official domain.}
        \label{fig:our-fp1}
    \end{subfigure}\hfill
    \begin{subfigure}[t]{0.48\linewidth}
        \centering
        \fbox{\includegraphics[width=\linewidth,height=4cm]{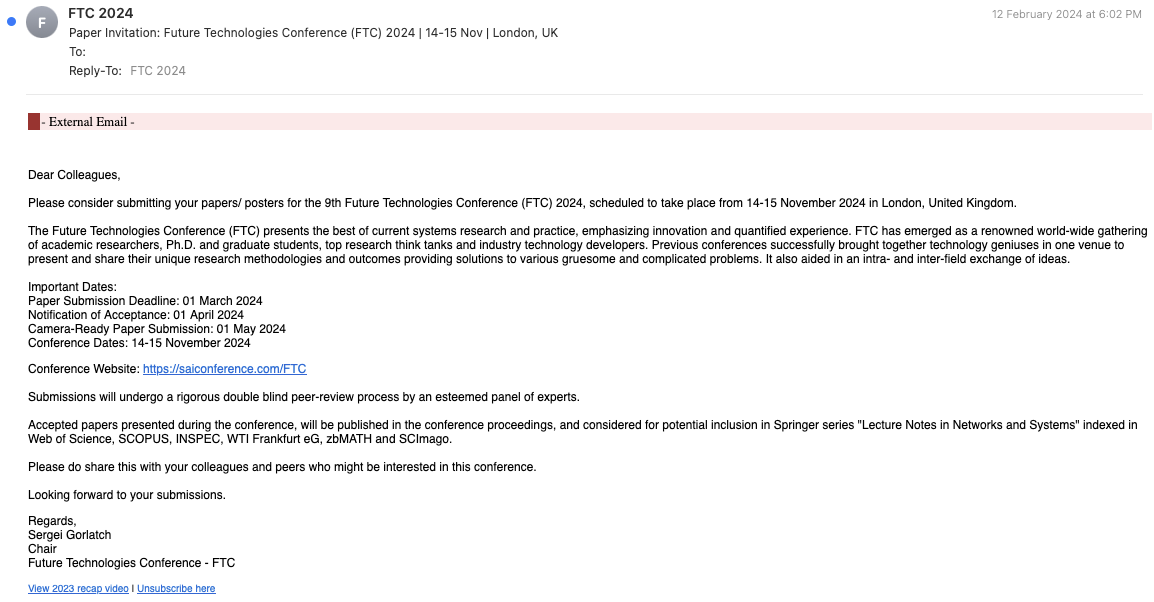}}
    \caption{\textbf{FP2} Notification claiming the Future
    Technologies Conference, sent from a personal Gmail account
    rather than the organizing committee's official domain.}
        \label{fig:our-fp2}
    \end{subfigure}
\vspace{1pt}
    \begin{subfigure}[t]{0.48\linewidth}
        \centering
        \fbox{\includegraphics[width=\linewidth,height=5cm]{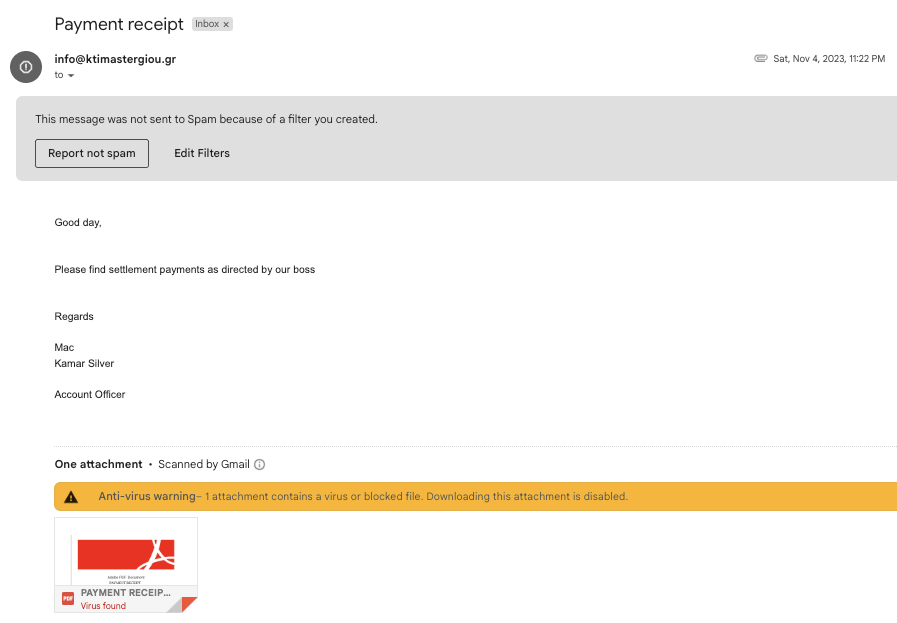}}
        \caption{\textbf{FN1} Impersonates a co-worker and delivers a malicious attachment. The email contains no link or explicit call-to-action phrase, so the instruction recognition module does not fire.}
        \label{fig:our-fn1}
    \end{subfigure}\hfill
    \begin{subfigure}[t]{0.48\linewidth}
        \centering
        \fbox{\includegraphics[width=\linewidth,height=5
        cm]{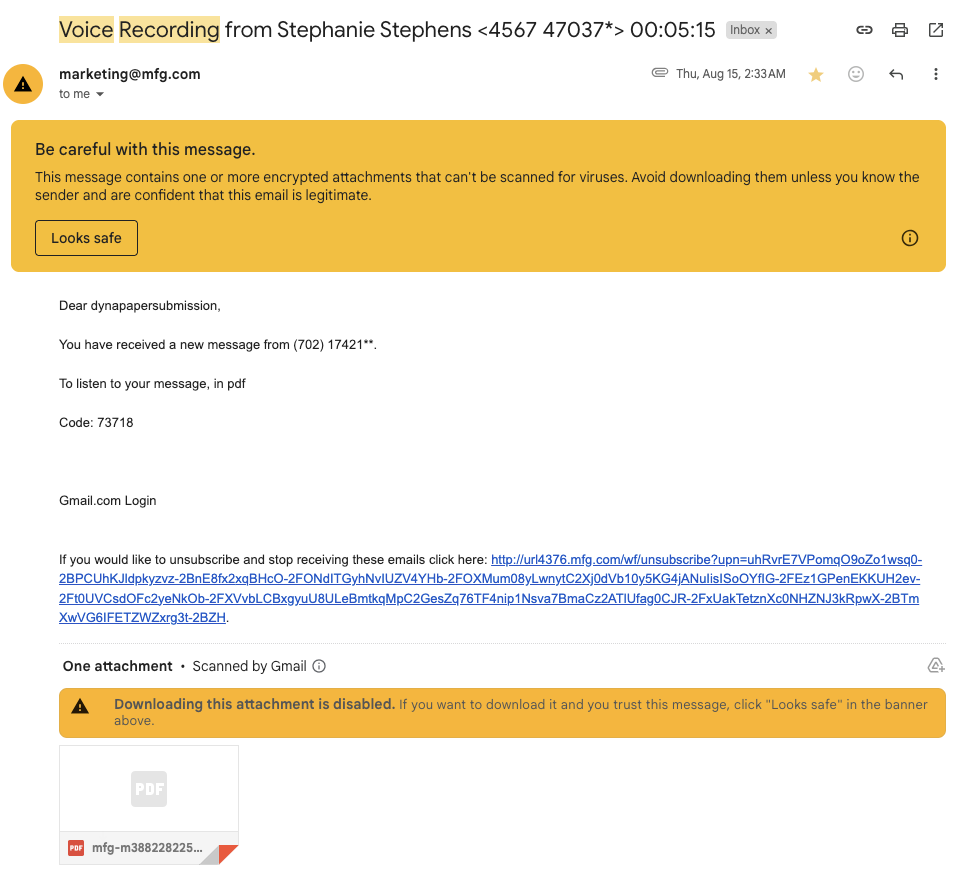}}
    \caption{\textbf{FN2} Claims an ambiguous sender identity that
    cannot be resolved against the knowledge base, so no
    identity-domain contradiction can be established; the payload is again a malicious attachment.}
        \label{fig:our-fn2}
    \end{subfigure}
    \caption{Failure cases in the open-world experiment.}
    \label{fig:openworld-failure-1}
\end{figure*}

\end{document}